# Nonlinear response functions and disorder: The case of the photogalvanic effect

Konstantinos Ladovrechis * and Tobias Meng

*Institute of Theoretical Physics and Würzburg-Dresden Cluster of Excellence ct.qmat, Technische Universität Dresden, 01062 Dresden, Germany*

(Received 5 October 2023; accepted 19 August 2025; published 29 September 2025)

We investigate the impact of disorder in the form of impurity scattering on a generalized version of the circular photogalvanic effect (CPGE) in Weyl semimetals where the frequency detuning between the two orthogonally polarized beams is nonzero. Considering a minimal model with two Weyl nodes at different energies, we employ the self-consistent Born approximation to unravel the dependence of the associated two-point retarded Green's function on the strength of intra- and internode scattering, frequency detuning, and energy difference between the two Weyl nodes. In the case of intranode scattering only, the second-order current density acquires Drude-like features, which we elucidate further by introducing an effective scattering strength. The Drude-like theory can even describe the second-order response in the presence of strong internode scattering, provided the latter has a linear interdependence with the intranode scattering. By properly adjusting the frequency detuning, we also find the real part of the two-point retarded Green's function to be reminiscent of a "quantized CPGE"-like form, although the imaginary part of the latter function is in general finite, and the second-order current density oscillates with time due to the finite frequency detuning. We finally conclude with an outlook on possible experimental consequences.

## I. INTRODUCTION

The optical response of solid-state systems has long been an attractive area of research, for example, with regard to second-order photovoltaic effects [1–3]. A particularly interesting case is the circular photogalvanic effect (CPGE), which is the spontaneous generation of a zero-frequency current due to the interference among one-photon absorption processes arising from two orthogonally polarized light beams of opposite frequency [4]. The occurence of CPGE was first predicted and experimentally demonstrated in the semiconductor tellurium [5–7]. Initially, the CPGE has been formulated in terms of the density matrix expressed as a time-independent perturbation series with respect to the interaction between electrons and photons [8,9]. However, such an approach led to the prediction of a divergent DC current density. The latter issue was resolved in the reformulation of the CPGE within the nonlinear susceptibility framework by the proper consideration of sum rules [10,11]. That development was followed by experimental observations of the CPGE taking place in p-GaAs/AlGaAs multiple-quantum wells [12,13], Si-metal-oxide-semiconductor transistors [14], and transition-metal dichalcogenides [15]. Concurrently, *ab initio* calculations predicted the existence of the CPGE in single-layer germanium and tin monochalcogenides [16], while its dependence on excitonic and correlation effects has been determined for the ferroelectric insulator $BaTiO_3$ and the monochalcogenide compound SnSe [17].

The advent of topological materials [18–21] invigorated the interest in the CPGE, especially when the latter was rederived in terms of the Berry connection using the Floquet formalism [22]. The application of nonequilibrium Green's function methods successfully addressed the corresponding temporal response [23–26], while the use of the Boltzmann equation provided insights into the CPGE response under spatially inhomogeneous light illumination [27]. Subsequent works connected the CPGE to the Fermi-liquid property of anomalous velocity caused by the orbital Berry phase of Bloch electrons [28], as well as to the intrinsic anomalous nonlinear Hall response in three-dimensional materials characterized by nonvanishing Berry-curvature dipole [29].

In particular, for Weyl semimetals, the generation rate of the CPGE current density of a photo-active untilted Weyl cone has been suggested to be quantized (but not topologically protected) in terms of fundamental constants in the absence of interactions [30,31]. Signs of a quantized CPGE response have been recently observed in the Weyl semimetal compound CoSi [32]. Other proposals have contemplated nonuniversal features, such as the dependence of the CPGE on the finite tilt of the Weyl cone and its topological charge [33–35]. Furthermore, it has also been suggested that a nonvanishing CPGE injection rate is accompanied by the generation of an effective chiral magnetic effect leading to a giant current-density response through spin-momentum locking at low frequencies [36], while the presence of an emergent electromagnetic induction in momentum space may enhance that injection rate

*Contact author: kladovrechis@gmail.com

[37]. In addition, the recent observation of CPGE in TaAs and RhSi [38–42], as well as the subsequent generation of spatially dispersive CPGE in $MoSe_2$ [43] amplified the theoretical and experimental interest toward investigating the link between CPGE and benchmark transport properties of Weyl semimetals. In particular, the Fermi arc surface states [44,45] have been investigated as a means to enable and control the confinement-induced CPGE via the characteristics of the boundary [46,47].

In this paper, we analyze the fate of the second-order optical response in Weyl semimetals, part of which is the CPGE, in the presence of impurity scattering when the two orthogonally polarized light beams have a finite frequency detuning. To that end, we focus on a minimal system with two Weyl nodes located at different energies. The finite detuning is considered adequately smaller compared to both optical frequencies. Introducing disorder in the form of elastic, pointlike intra- and internode scattering, we study the impact of impurities on the generalized second-order optical response by determining the behavior of the real and imaginary parts of the corresponding correlation function in terms of the intra- and internode scattering strengths, frequency detuning, and energy difference between the two Weyl nodes. In the case of intranode scattering, the optical response is reminiscent of the Drude-like behavior of the AC linear optical conductivity in metals [48]. We further characterize such a response through the introduction of a simple effective model with an effective scattering time. The applicability of the effective model is discussed as a function of the external frequency of the light beams, the frequency detuning, and the intranode scattering strength. Upon including internode scattering as well, the simple model can describe the data to a limited degree, but only when inter- and intranode scattering are assumed to be linearly dependent to one another. The remainder of the paper is organized as follows. In Sec. II, the model Hamiltonian for the Weyl semimetal is introduced and analyzed, while the formalism for the second-order optical response is reviewed in Sec. III. The treatment of disorder on the level of a self-consistent Born approximation (SCBA) is detailed in Sec. IV. The effective Drude-like theory for intranode scattering is derived and discussed in Sec. V, and it is benchmarked by a direct comparison to the fully numerical analysis in Sec. VI. The impact of internode scattering is detailed in Sec. VII. The results are finally summarized in Sec. VIII, where we also comment on experimental prospects.

## II. THE MODEL

We focus on the low-energy regime of a time-reversal-symmetry-broken and noncentrosymmetric Weyl semimetal with the minimal number of two Weyl pockets labeled by their chirality $\chi = \pm 1$. Both pockets are centered around an isotropic Weyl node. Such a setting can be modeled by a momentum-linearized two-band Hamiltonian operator $\hat{H}_{\text{Weyl}} = \sum_\chi \hat{H}_\chi$ with the Weyl node with chirality $\chi$ to be described by the following operator:

$$\hat{H}_\chi = \sum_{\boldsymbol{p}}{}' \hat{\Psi}^\dagger_{\boldsymbol{p}} \left(\chi\, v_F\, \boldsymbol{\sigma} \cdot (\boldsymbol{p} - \boldsymbol{p}_\chi) - \mu_\chi\right) \hat{\Psi}_{\boldsymbol{p}}. \quad (1)$$

Parameter $v_F$ denotes the Fermi velocity (taken to be identical at both nodes for simplicity), $\boldsymbol{\sigma} = \{\sigma_1, \sigma_2, \sigma_3\}$ corresponds to the vector of Pauli matrices, $\boldsymbol{p}_\chi$ is the momentum of the Weyl node with chirality $\chi$, and $\mu_\chi$ is the *energy offset*, i.e., the energy difference between that node and the chemical potential. The fermionic spinor is defined as $\hat{\Psi}^\dagger_{\boldsymbol{p}} = (\hat{c}^\dagger_{\boldsymbol{p}\uparrow}, \hat{c}^\dagger_{\boldsymbol{p}\downarrow})$, where the operator $\hat{c}^\dagger_{\boldsymbol{p}s}$ creates an electron of spin $s = \{\uparrow, \downarrow\}$ and three-dimensional momentum $\boldsymbol{p}$. The prime index in Eq. (1) indicates that the model is an effective low-energy theory, and it comes with the momentum cutoff $\Lambda/v_F \geqslant |\boldsymbol{p} - \boldsymbol{p}_\chi|$ that is much smaller than the separation of the two Weyl nodes in momentum space.

Thus far in Eq. (1), the momentum $\boldsymbol{p}$ and the associated fermionic spinors $\{\hat{\Psi}^\dagger_{\boldsymbol{p}}, \hat{\Psi}_{\boldsymbol{p}}\}$ have been defined in the entire Brillouin zone. However, since the applied electric field is spatially homogeneous and only the zero-momentum component of the two-point retarded Green's function contributes to the quadratic response (see Sec. III), the current-density operators characterizing the Weyl semimetal are hence pocket-local. It is therefore convenient to introduce the pocket-specific notation $\hat{\Psi}_{\chi\boldsymbol{k}} = \hat{\Psi}_{\boldsymbol{p}_\chi + \boldsymbol{k}}$, where $\boldsymbol{k}$ is the momentum relative to Weyl node with chirality $\chi$. Then, the Hamiltonian operator in Eq. (1) reads

$$\hat{H}_\chi = \sum_{\boldsymbol{k}}{}' \hat{\Psi}^\dagger_{\chi\boldsymbol{k}} \left(\chi\, v_F\, \boldsymbol{\sigma} \cdot \boldsymbol{k} - \mu_\chi\right) \hat{\Psi}_{\chi\boldsymbol{k}}. \quad (2)$$

In order to allow for a generic situation in which intra- and internode impurity scattering can take place, we introduce the operator

$$\hat{H}_{\text{imp}} = \sum_{\chi,\chi'=\pm 1} \int d^3r\, \hat{\Psi}^\dagger_\chi(\boldsymbol{r}) \sum_{j=1}^{N_{\text{imp}}} V_{\chi\chi'}(\boldsymbol{r} - \boldsymbol{R}_j)\, \hat{\Psi}_{\chi'}(\boldsymbol{r}), \quad (3)$$

where $\hat{\Psi}^\dagger_\chi(\boldsymbol{r})$ is the real-space Fourier transform of the pocket-specific spinor $\hat{\Psi}^\dagger_{\chi\boldsymbol{k}}$. Parameter $N_{\text{imp}}$ denotes the total number of randomly distributed impurity centres. The positions of the impurity centres are represented by the vector $\boldsymbol{R}_j$, and the disorder potential is chosen as [49]

$$V_{\chi\chi'}(\boldsymbol{r} - \boldsymbol{R}_j) = V_{0,\chi\chi'}\, \mathbb{1}_2\, \delta(\boldsymbol{r} - \boldsymbol{R}_j). \quad (4)$$

The latter represents an effectively pointlike elastic scattering event, mediated by the $j$th impurity center, among electronic states around a Weyl node with chirality $\chi$ and a Weyl node with chirality $\chi'$. The parameter $V_{0,\chi\chi'}$ denotes the amplitude of the aforementioned scattering process. Therefore, the complete model for the disordered Weyl semimetal is represented by the following operator:

$$\hat{H}_0 = \sum_{\chi=\pm 1} \hat{H}_\chi + \hat{H}_{\text{imp}}. \quad (5)$$

## III. QUADRATIC RESPONSE AND ITS DISORDER-FREE LIMIT

We consider the Weyl semimetal to be subjected to an electric field that induces a nonequilibrium state. While restricting the electric field to be spatially homogeneous, we allow, for the time being, the latter to be a superposition of contributions at different optical frequencies, i.e., $\boldsymbol{E}(t) = \sum_{n=1}^{2} \boldsymbol{E}_n\, e^{-i\omega_n t}$.

The constant coefficients $\boldsymbol{E}_n$ must be chosen such that the total field is real. In the remaining part of the current paper, we focus on the quadratic response of the Weyl semimetal in the presence of the applied electric field $\boldsymbol{E}(t)$. Such a situation assumes that the latter represents a *sufficiently small* perturbation away from equilibrium state, described by Eq. (5), that can be treated using a perturbative scheme where the expansion parameter is the amplitude of the applied electric field. As detailed in Appendix A, the $\alpha$th component of the current-density operator, parametrized in real space-time coordinates as $\hat{j}_\alpha(\boldsymbol{r}, t)$, contains a part that is of second order in the applied electric field $\boldsymbol{E}(t)$. That part gives rise to the second-order current density

$$\langle \hat{j}_\alpha(\boldsymbol{r}, t)\rangle^{(\mathrm{quad})} = \sum_{n,m=1}^{2} \frac{e^{i(\omega_n+\omega_m)t}}{-\,\omega_n\,\omega_m} \times \sum_{\beta,\gamma} \mathcal{X}_{\alpha\beta\gamma}(\mathbf{0},\mathbf{0};\omega_n,\omega_m)\, E_{n,\beta} E_{m,\gamma}. \tag{6}$$

The Greek indices $\{\alpha, \beta, \gamma\}$ run through the spatial coordinates $\{x, y, z\}$, and the coefficient $E_{\{1,2\},\beta}$ ($E_{\{1,2\},\gamma}$) is the $\beta$th ($\gamma$th) component of a particular electric-field mode. In Fourier domain, the two-point retarded Green's function $\mathcal{X}_{\alpha\beta\gamma}(\mathbf{0},\mathbf{0};\omega_1,\omega_2)$ has zero entries for the corresponding momenta, since the application of a spatially homogeneous electric field $\boldsymbol{E}(t)$ prompts only the zero-momentum mode of the response function to be excited. $\mathcal{X}_{\alpha\beta\gamma}(\mathbf{0},\mathbf{0};\omega_1,\omega_2)$ is connected to the imaginary-time three-current correlation function $\widetilde{\mathcal{X}}_{\alpha\beta\gamma}(\mathbf{0},\mathbf{0};i\Omega_1,i\Omega_2)$ via the relation

$$\mathcal{X}_{\alpha\beta\gamma}(\mathbf{0},\mathbf{0};\omega_1,\omega_2) = \frac{\widetilde{\mathcal{X}}_{\alpha\beta\gamma}(\mathbf{0},\mathbf{0};i\Omega_1,i\Omega_2)|_{i\Omega_{1,2}\to\omega_{1,2}+i\,\eta_{1,2}}}{2}. \tag{7}$$

In the latter expression, $\Omega_{1,2}$ denote bosonic Matsubara frequencies, while parameters $\eta_{1,2}$ are auxiliary infinitesimal factors. The imaginary-time three-current correlation function is computed through the expression

$$\begin{aligned}\widetilde{\mathcal{X}}_{\alpha\beta\gamma}(\mathbf{0},\mathbf{0};i\Omega_1,i\Omega_2) &= \int d^3r_1 \int_0^\beta d\tau_1 \int d^3r_2 \int_0^\beta d\tau_2 e^{i\Omega_1\tau_1} e^{i\Omega_2\tau_2} \\ &\quad\times \langle T_\tau\, \hat{J}_\alpha(0,0)\hat{J}_\beta(-\boldsymbol{r}_1,-\tau_1)\hat{J}_\gamma(-\boldsymbol{r}_2,-\tau_2)\rangle,\end{aligned} \tag{8}$$

where $\hat{J}_\alpha(\boldsymbol{r},\tau)$ is the $\alpha$th component of the current-density operator in imaginary-time formalism. In the pocket-specific formalism of Sec. III, the $\alpha$th component of the current-density operator in the vicinity of the Weyl node with chirality $\chi$ reads

$$\hat{J}_\alpha^{(\chi)} = {\sum_{\boldsymbol{k}}}' \hat{\Psi}^\dagger_{\chi\boldsymbol{k}}\, J_{0,\alpha}^{(\chi)}\, \hat{\Psi}_{\chi\boldsymbol{k}}, \tag{9}$$

where the corresponding component of the vertex function is given by

$$J_{0,\alpha}^{(\chi)} = \chi\, v_F\,(-e)\,\sigma_\alpha, \tag{10}$$

with $-e < 0$ being the electron charge. Hence, the three-current correlation function in Eq. (8) is also pocket-local. It can thus be written as the sum of contributions from both pockets in the following form [50]:

$$\widetilde{\mathcal{X}}_{\alpha\beta\gamma}(i\Omega_1,i\Omega_2) = \widetilde{\mathcal{X}}^{(+1)}_{\alpha\beta\gamma}(i\Omega_1,i\Omega_2) + \widetilde{\mathcal{X}}^{(-1)}_{\alpha\beta\gamma}(i\Omega_1,i\Omega_2), \tag{11}$$

where we have suppressed the zero-momentum entries for simplicity. Using Eq. (10) and the bare Matsubara Green's function with momentum $\boldsymbol{q}$ and fermionic Matsubara frequency $\omega_n$ at the Weyl node with chirality $\chi$,

$$G^{(\chi,0)}_{\boldsymbol{q},\omega_n} = \sum_{b=\pm1} \frac{\frac{1}{2}(\mathbb{1} + b\,\boldsymbol{\sigma}\cdot\boldsymbol{q})}{i\omega_n - b\,\chi\,v_F\,q + \mu_\chi}, \tag{12}$$

the three-current correlation function at the Weyl node with chirality $\chi$ reads

$$\begin{aligned}&\widetilde{\mathcal{X}}^{(\chi)}_{\alpha\beta\gamma}(i\Omega_1,i\Omega_2) \\ &= \frac{1}{T^{-1}V}\sum_{\boldsymbol{q},\omega_n}\Big[\mathrm{Tr}\Big(G^{(\chi,0)}_{\boldsymbol{q},\Omega_1+\Omega_2+\omega_n}\,J^{(\chi)}_{0,\beta}\,G^{(\chi,0)}_{\boldsymbol{q},\Omega_2+\omega_n}\,J^{(\chi)}_{0,\gamma}\,G^{(\chi,0)}_{\boldsymbol{q},\omega_n}\,J^{(\chi)}_{0,\alpha}\Big) \\ &\quad + (\beta\leftrightarrow\gamma,\ \Omega_1\leftrightarrow\Omega_2)\Big],\end{aligned} \tag{13}$$

where $T$ is the temperature, $V$ denotes the system volume, while the last term in the rectangular bracket has the same form as the first one with the indices swapped as indicated. Conveniently, the physical content of Eq. (13) is interpreted via the triangle diagrams shown in Fig. 1: Since the three-current correlation function $\widetilde{\mathcal{X}}^{(\chi)}_{\alpha\beta\gamma}(i\Omega_1,i\Omega_2)$ describes the generation of a current density (vertex $\alpha$) after two consecutive interactions between electrons and photons (vertices $\beta$ and $\gamma$), the two triangle diagrams correspond to the two distinct relative time orderings of the one-photon absorption events.

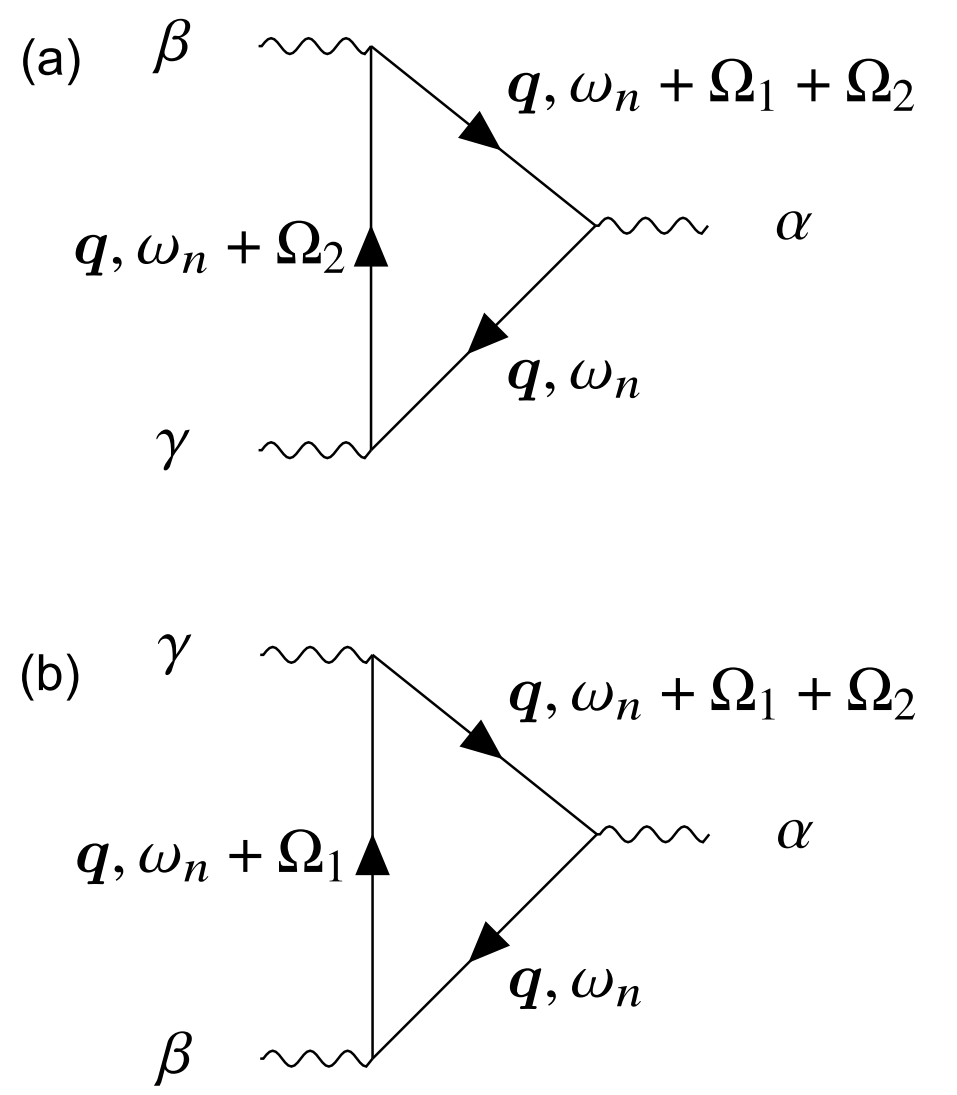

FIG. 1. Topologically distinct diagrams contributing to the correlation function $\widetilde{\mathcal{X}}^{(\chi)}_{\alpha\beta\gamma}(i\Omega_1, i\Omega_2)$ given in Eq. (13).

## IV. SELF-CONSISTENT BORN APPROXIMATION

The impact of disorder on the CPGE is to be treated perturbatively in terms of the strength of impurity scattering represented by the scattering amplitude $V_{0,\chi\chi'}$. For such a purpose, we are going to implement an analysis on the level of the self-consistent Born approximation. The latter is defined as the resummation of diagrams corresponding to fermionic self-energy and vertex corrections of the types shown in Fig. 2. Those diagrams are quadratic functions of the scattering amplitude ($\propto V_{0,\chi\chi'}V_{0,\chi'\chi''}$). We note that, within the SCBA scheme, the self-energy correction due to tadpole diagrams is neglected, since its contribution can always be absorbed into the definition of the energy offset at every Weyl node. To naturally incorporate the impact of impurity scattering on the CPGE via the diagrams shown in Fig. 2, we generalize the three-current correlation function given in Eq. (13) to an expression featuring *renormalized* Matsubara Green's functions and vertex functions. However, we stress that there are several scattering processes which *can* occur in Eq. (13), but they are neglected. Examples of such processes are shown in Fig. 3. Diagram (a) describes a triple scattering event of an electron from a single impurity. The corresponding self-energy correction is proportional to $V_{0,\chi\chi'}V_{0,\chi'\chi''}V_{0,\chi''\chi'''}$ and it lies beyond that of the considered order $V_{0,\chi\chi'}V_{0,\chi'\chi''}$. Diagram (b) is a two-electron scattering event from two impurity centres. It is associated with a self-energy "crossing" term that has a significantly smaller available phase space for the scattering vectors compared to the self-energy correction in Fig. 2. Thus, that type of diagrams is neglected. Diagram (c) describes three-electron scattering from a single impurity, whereas diagram (d) is a two-electron scattering event, where one of the electrons has its energy difference from the chemical potential renormalized. In those two cases, the corrections cannot be classified as either purely self-energy or purely vertex ones within the SCBA resummation scheme, thus they are neglected.

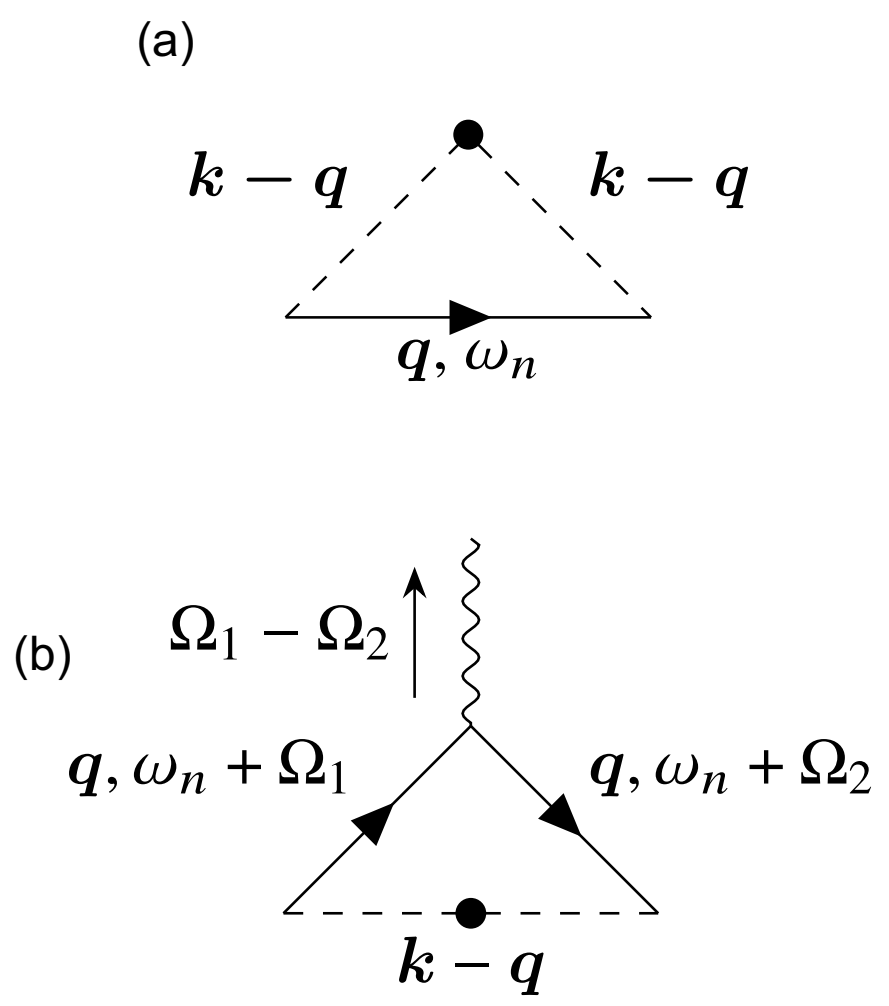

FIG. 2. One-particle irreducible diagrams of order $V_{0,\chi\chi'}V_{0,\chi'\chi''}$ representing (a) the self-energy correction and (b) the vertex correction.

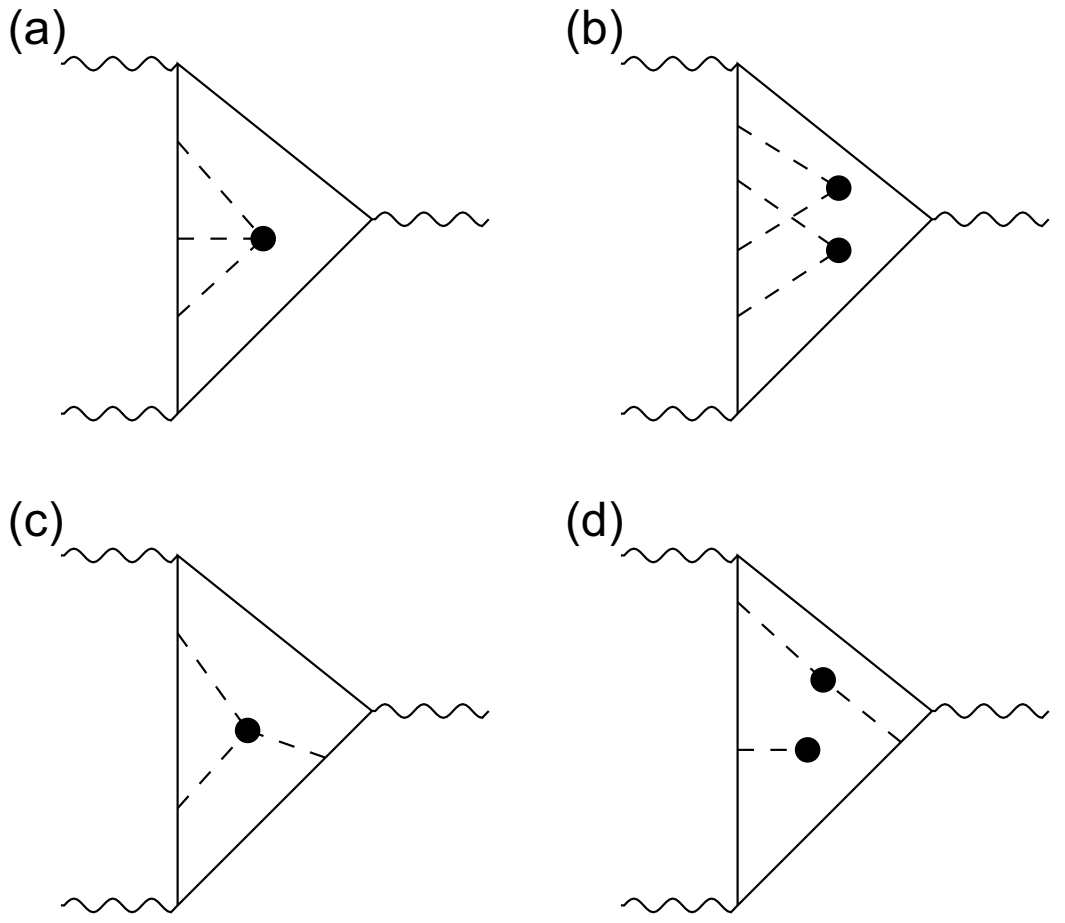

FIG. 3. Examples of scattering events occurring in the three-current correlation function [see Eq. (13) that cannot be captured under the SCBA scheme (self-energy and vertex corrections shown in Fig. 2).

Replacing the scattering amplitude $V_{0,\chi\chi'}$ in Eq. (3) by the value $V_{\text{intra}}$ when $\chi = \chi'$, i.e., for electronic states scattered within the same pocket (*intranode scattering*), and by the value $V_{\text{inter}}$ when $\chi \neq \chi'$, i.e., for electronic states scattered in-between different pockets (*internode scattering*), the bare Matsubara Green's function in Eq. (12) is then renormalized as

$$G^{(\chi,\text{SCBA})}_{\boldsymbol{q},\omega_n} = \sum_{b=\pm 1} \frac{\frac{1}{2}(\mathbb{1} + b\,\boldsymbol{\sigma}\cdot\boldsymbol{q})}{i\omega_n - b\,\chi\, v_F\, q + \mu_\chi - \Sigma^{(\chi)}_{\text{SCBA}}(i\omega_n)}. \quad (14)$$

The self-energy correction $\Sigma^{(\chi)}_{\text{SCBA}}(i\omega_n)$ is the solution of the following self-consistent equation:

$$\begin{aligned}\Sigma^{(\chi)}_{\text{SCBA}}(i\omega_n) = {} & \widetilde{W}_{\text{intra}}\left(i\omega_n + \mu_\chi - \Sigma^{(\chi)}_{\text{SCBA}}(i\omega_n)\right) \\ & \times \Bigg[-1 + \frac{i\omega_n + \mu_\chi - \Sigma^{(\chi)}_{\text{SCBA}}(i\omega_n)}{2 v_F \Lambda} \\ & \times \ln\left(\frac{i\omega_n + \mu_\chi - \Sigma^{(\chi)}_{\text{SCBA}}(i\omega_n) + \Lambda}{i\omega_n + \mu_\chi - \Sigma^{(\chi)}_{\text{SCBA}}(i\omega_n) - \Lambda}\right)\Bigg] \\ & + (\widetilde{W}_{\text{intra}} \to \widetilde{W}_{\text{inter}},\ \mu_\chi \to \mu_{-\chi}). \end{aligned} \quad (15)$$

The last term in Eq. (15) has the same form as the first one with the parameters changed as indicated. Parameter $\rho_{\text{imp}}$ denotes the density of impurity centres, while the *dimensionless* intranode (internode) scattering strength $\widetilde{W}_{\text{intra (inter)}}$ is defined as

$$\widetilde{W}_{\text{intra (inter)}} := \frac{|V_{\text{intra (inter)}}|^2\, \rho_{\text{imp}}\, \Lambda}{2\pi^2 v_F^3}. \quad (16)$$

Regarding the renormalization of the vertex function present in Eq. (13) and connecting a Matsubara Green's function at frequency $i\omega_n$ to a Matsubara Green's function at $i(\omega_n + \Omega)$, we arrive at the following expression:

$$\boldsymbol{J}^{(\chi)}_{\text{SCBA}}(i\omega_n, i\Omega) = \frac{\chi\, v_F\, (-e)}{1 - Y^{(\chi)}_{\text{SCBA}}(i\omega_n, i\Omega)}\boldsymbol{\sigma}. \quad (17)$$

The vertex correction is given by

$$Y^{(\chi)}_{\text{SCBA}}(i\omega_n, i\Omega) = \frac{1}{6\pi^2(\chi v_F)^3}\Bigg[-2\Lambda + \ln\left(\frac{i\omega_n + \mu_\chi - \Sigma^{(\chi)}_{\text{SCBA}}(i\omega_n) + \Lambda}{i\omega_n + \mu_\chi - \Sigma^{(\chi)}_{\text{SCBA}}(i\omega_n) - \Lambda}\right)\left(i\omega_n + \mu_\chi - \Sigma^{(\chi)}_{\text{SCBA}}(i\omega_n)\right)^2 \times\left(-\frac{1}{i\omega_n + \mu_\chi - \Sigma^{(\chi)}_{\text{SCBA}}(i\omega_n) - i\Omega + \Sigma^{(\chi)}_{\text{SCBA}}(i\Omega)} + \frac{2}{i\omega_n + \mu_\chi + i\Omega - \Sigma^{(\chi)}_{\text{SCBA}}(i\omega_n) - \Sigma^{(\chi)}_{\text{SCBA}}(i\Omega)}\right) + (\omega_n \leftrightarrow \Omega)\Bigg]. \tag{18}$$

The last term in the rectangular brackets in Eq. (18) has the same form as the first one with the frequencies swapped as indicated.

Henceforth, we focus on the regime for the fermionic energy $\omega$ about a Weyl node with chirality $\chi$ determined the following two conditions [51]. First, that through the SCBA resummation scheme the considered energy regime is sufficiently smaller than the energy cutoff $\Lambda$:

$$\left|\omega + i0^+ - \text{Re}\Sigma^{(\chi)}_{\text{SCBA}}(\omega + i0^+)\right| \ll \Lambda. \tag{19}$$

Second, that the broadening of the spectral density function is sufficiently smaller as well compared to the energy cutoff:

$$\left|\text{Im}\Sigma^{(\chi)}_{\text{SCBA}}(\omega + i0^+\right| \ll \Lambda. \tag{20}$$

Under the aforementioned two inequalities, the self-consistent Eq. (15) can be approximately but analytically solved to yield

$$\Sigma^{(\chi)}_{\text{SCBA, approx}}(\omega) = \omega + i0^+ + \frac{\mu_\chi \widetilde{W}_{\text{intra}} + \mu_{-\chi}\widetilde{W}_{\text{inter}}}{\widetilde{W}_{\text{intra}} + \widetilde{W}_{\text{inter}}} + i\Lambda\frac{1 - \widetilde{W}_{\text{intra}} - \widetilde{W}_{\text{inter}}}{\pi(\widetilde{W}_{\text{intra}} + \widetilde{W}_{\text{inter}})} - i\frac{\Lambda}{\pi(\widetilde{W}_{\text{intra}} + \widetilde{W}_{\text{inter}})} \times\left[(1 - \widetilde{W}_{\text{inter}} - \widetilde{W}_{\text{intra}})^2 + \frac{\widetilde{W}_{\text{intra}}\widetilde{W}_{\text{inter}}\pi^2(\mu_\chi - \mu_{-\chi})^2}{\Lambda^2} - i\frac{2\pi\Lambda(\mu_\chi\widetilde{W}_{\text{intra}} - \mu_{-\chi}\widetilde{W}_{\text{inter}} + \omega(\widetilde{W}_{\text{intra}} + \widetilde{W}_{\text{inter}}))}{(v_F\Lambda)^2}\right]^{1/2}, \tag{21}$$

where the analytic continuation $i\omega_n \to \omega + i0^+$ has been performed. The perturbative behavior of Eq. (15) as well as Eq. (21) is controlled by the parameter [51]

$$\frac{(1 - \widetilde{W}_{\text{inter}} - \widetilde{W}_{\text{intra}})^2\Lambda}{|\omega|}. \tag{22}$$

In the absence of internode scattering while $\widetilde{W}_{\text{intra}} = 1$, we thus find that the zero-frequency limit of the following expression is divergent:

$$\lim_{\omega\to 0}\frac{\text{Re}\Sigma^{(\chi)}_{\text{SCBA, approx}}(\omega)}{|\omega|}\Bigg|_{\widetilde{W}_{\text{inter}}=0} \longrightarrow -\infty. \tag{23}$$

Therefore, the value $\widetilde{W}_{\text{intra,crit}} = 1$ is termed *critical scattering strength*, and it is the upper threshold for the interval of values for $\widetilde{W}_{\text{intra}}$ considered in the rest of the current paper. Using Eq. (21) as an input to Eqs. (14), (17), and (18), we then generalize Eq. (13) to the following *renormalized* three-current correlation function under the SCBA scheme:

$$\widetilde{\mathcal{X}}^{(\chi,\text{SCBA})}_{\alpha\beta\gamma}(i\Omega_1, i\Omega_2) = \frac{1}{T^{-1}V}\sum_{\boldsymbol{q},\omega_n}\Big[\text{Tr}\Big(G^{(\chi,\text{SCBA})}_{\boldsymbol{q},\Omega_1+\Omega_2+\omega_n} J^{(\chi)}_{\text{SCBA},\beta}(i\omega_n + \Omega_2, i\Omega_1)\, G^{(\chi,\text{SCBA})}_{\boldsymbol{q},\Omega_2+\omega_n} \times J^{(\chi)}_{\text{SCBA},\gamma}(i\omega_n, i\Omega_2)\, G^{(\chi,\text{SCBA})}_{\boldsymbol{q},\omega_n} J^{(\chi)}_{\text{SCBA},\alpha}(i\omega_n, i\Omega_1 + \Omega_2)\Big) + (\beta \leftrightarrow \gamma, \Omega_1 \leftrightarrow \Omega_2)\Big]. \tag{24}$$

We note that the presence of logarithmic terms in the vertex corrections [see Eq. (18)] gives rise to branch cuts along the combinations of bosonic Matsubara frequencies present in Eq. (24). Therefore, deforming the discrete summation over the fermionic Matsubara frequencies $\omega_n$ into integrals running along the axis of the *real* fermionic frequency $\omega$, we obtain the following analytical expression for the three-current correlation function:

$$
\begin{aligned}
&\widetilde{\mathcal{X}}^{(\chi,\mathrm{SCBA})}_{\alpha\beta\gamma}(i\Omega_1, i\Omega_2)\\
&= \frac{i}{3}\chi\epsilon^{\alpha\beta\gamma}(-e)^3 v_F^3 \int_0^{\Lambda/v_F} dk\frac{k^2}{2\pi^2}\sum_{b_1,b_2,b_3=\pm1}(1-\delta_{b_1b_2}\delta_{b_1b_3})\int_{-\infty}^{\infty} d\omega\, n_F(\omega)\\
&\times\Bigg(\Bigg[\frac{1}{1-Y^{(\chi)}_{\mathrm{SCBA}}(\omega+i\eta,\omega+i\Omega_1+i\Omega_2)}\frac{1}{1-Y^{(\chi)}_{\mathrm{SCBA}}(\omega+i\eta+i\Omega_2,\omega+i\eta)}\frac{1}{\omega+i\eta+\mu_\chi-b_3v_Fk-\Sigma^{(\chi)}_{\mathrm{SCBA}}(\omega+i\eta)}\\
&-\frac{1}{1-Y^{(\chi)}_{\mathrm{SCBA}}(\omega-i\eta,\omega+i\Omega_1+i\Omega_2)}\frac{1}{1-Y^{(\chi)}_{\mathrm{SCBA}}(\omega-i\eta+i\Omega_2,\omega-i\eta)}\frac{1}{\omega-i\eta+\mu_\chi-b_3v_Fk-\Sigma^{(\chi)}_{\mathrm{SCBA}}(\omega-i\eta)}\Bigg]\\
&\times\frac{1}{\omega+\mu_\chi+\Omega_1+i\Omega_2-b_1v_Fk-\Sigma^{(\chi)}_{\mathrm{SCBA}}(\omega+i\Omega_1+i\Omega_2)}\frac{1}{\omega+\mu_\chi+i\Omega_2-b_2v_Fk-\Sigma^{(\chi)}_{\mathrm{SCBA}}(\omega+i\Omega_2)}\\
&\times\frac{1}{1-Y^{(\chi)}_{\mathrm{SCBA}}(\omega+i\Omega_1+i\Omega_2,\omega+i\Omega_2)}\\
&+\Bigg[\frac{1}{1-Y^{(\chi)}_{\mathrm{SCBA}}(\omega+i\Omega_1,\omega+i\eta)}\frac{1}{1-Y^{(\chi)}_{\mathrm{SCBA}}(\omega+i\eta,\omega-i\Omega_2)}\frac{1}{\omega+i\eta+\mu_\chi-b_2v_Fk-\Sigma^{(\chi)}_{\mathrm{SCBA}}(\omega+i\eta)}\\
&-\frac{1}{1-Y^{(\chi)}_{\mathrm{SCBA}}(\omega+i\Omega_1,\omega-i\eta)}\frac{1}{1-Y^{(\chi)}_{\mathrm{SCBA}}(\omega-i\eta,\omega-i\Omega_2)}\frac{1}{\omega-i\eta+\mu_\chi-b_2v_Fk-\Sigma^{(\chi)}_{\mathrm{SCBA}}(\omega-i\eta)}\Bigg]\\
&\times\frac{1}{\omega+\mu_\chi+i\Omega_1-b_1v_Fk-\Sigma^{(\chi)}_{\mathrm{SCBA}}(\omega+i\Omega_1)}\frac{1}{\omega+\mu_\chi-i\Omega_2-b_3v_Fk-\Sigma^{(\chi)}_{\mathrm{SCBA}}(\omega-i\Omega_2)}\\
&\times\frac{1}{1-Y^{(\chi)}_{\mathrm{SCBA}}(\omega-i\Omega_2,\omega+i\Omega_1)}\\
&+\Bigg[\frac{1}{1-Y^{(\chi)}_{\mathrm{SCBA}}(\omega+i\eta-i\Omega_1-i\Omega_2,\omega+i\eta)}\frac{1}{1-Y^{(\chi)}_{\mathrm{SCBA}}(\omega+i\eta,\omega-i\Omega_1)}\frac{1}{\omega+i\eta+\mu_\chi-b_1v_Fk-\Sigma^{(\chi)}_{\mathrm{SCBA}}(\omega+i\eta)}\\
&-\frac{1}{1-Y^{(\chi)}_{\mathrm{SCBA}}(\omega-i\eta-i\Omega_1-i\Omega_2,\omega-i\eta)}\frac{1}{1-Y^{(\chi)}_{\mathrm{SCBA}}(\omega-i\eta,\omega-i\Omega_1)}\frac{1}{\omega-i\eta+\mu_\chi-b_1v_Fk-\Sigma^{(\chi)}_{\mathrm{SCBA}}(\omega-i\eta)}\Bigg]\\
&\times\frac{1}{\omega+\mu_\chi-i\Omega_1-b_2v_Fk-\Sigma^{(\chi)}_{\mathrm{SCBA}}(\omega-i\Omega_1)}\frac{1}{\omega+\mu_\chi-i\Omega_1-i\Omega_2-b_3v_Fk-\Sigma^{(\chi)}_{\mathrm{SCBA}}(\omega-i\Omega_1-i\Omega_2)}\\
&\times\frac{1}{1-Y^{(\chi)}_{\mathrm{SCBA}}(\omega-i\Omega_1,\omega-i\Omega_1-i\Omega_2)}\Bigg)+(\beta\leftrightarrow\gamma,\Omega_1\leftrightarrow\Omega_2),
\end{aligned}\tag{25}
$$

where $n_F(\omega) = [1 + \exp(\omega/T)]^{-1}$ is the Fermi-Dirac distribution function.

## V. EFFECTIVE DRUDE-LIKE THEORY FOR INTRANODE SCATTERING

We begin by discussing the case of intranode scattering at a Weyl node of chirality $\chi$, Thus, we set $\widetilde{W}_{\mathrm{inter}} = 0$ in Eq. (25). To unearth the physical content of the three-current correlation function and to gain an analytical insight into the impact of impurity scattering on the second-order current density [see Eq. (6)], we perform several *ad hoc* approximations in Eq. (25); we will further comment about those approximations in Sec. VI. First, we neglect all vertex corrections by setting $Y^{(\chi)}_{\mathrm{SCBA}} = 0$. Second, the self-energy correction is substituted by the following momentum-independent level broadening:

$$\Sigma^{(\chi)}_{\mathrm{SCBA}}(z) \to -i\Gamma\mathrm{sgn}(\mathrm{Im}(z)),\tag{26}$$

where $z$ is a complex-valued frequency argument and $\Gamma^{-1}$ is a positive-valued parameter interpreted as an *effective* scattering time. Third, the spectral density functions, formed by the expressions inside the rectangular brackets in Eq. (25), are approximated by Dirac delta functions according to the prescription

$$\frac{1}{\pi}\frac{\Gamma}{(\omega - b_iv_Fk+\mu_\chi)^2+\Gamma^2} \to \delta(\omega - b_iv_Fk+\mu_\chi).\tag{27}$$

The index $i$ runs through the integers $\{1, 2, 3\}$. Such an approximation allows for the exact evaluation of all frequency integrals in Eq. (25). To make connection to the second-order current density in Eq. (6), we perform analytic continuation in Eq. (25) for the bosonic Matsubara frequencies, $i\Omega_{1,2} \to \omega_{1,2} + i\eta_{1,2}$ where $\omega_{1,2}$ are two positive-valued real frequencies, and $\eta_{1,2} \to 0^+$.

As detailed in Appendix B, the remaining momentum integrals in Eq. (25) are evaluated in the zero-temperature limit under the conditions $\Gamma \ll |\omega_1 + \omega_2|, |\omega_1|, |\omega_2|$. Since we are interested in the response to a superposition of two orthogonal and linearly polarized electric-field modes, we choose the real frequencies to be

$$\omega_1 = \omega_0, \quad \omega_2 = -\omega_0 + \delta\omega. \tag{28}$$

The positive-valued optical frequency and frequency detuning are defined as $\omega_0$ and $\delta\omega$, respectively. Through Eq. (7), the Drude-like two-point retarded Green's function at the Weyl node with chirality $\chi$ reads [see Eq. (B4)]

$$\mathcal{X}^{(\chi,\text{SCBA,Drude})}_{\alpha\beta\gamma}(\omega_0, -\omega_0 + \delta\omega) \approx \chi \frac{(-e)^3}{24\pi}\epsilon^{\alpha\beta\gamma} \frac{\omega_0^2}{\delta\omega + 2i\Gamma} \frac{1}{\pi} \text{Im}[\ln(4|\mu_\chi|^2 - (\omega_0 + i\Gamma)^2)]. \tag{29}$$

The total electric field must be elliptically polarized, thus reading

$$\boldsymbol{E}(t) = \begin{pmatrix} E_x \cos(\omega_1 t) \\ E_y \sin(\omega_2 t) \\ 0 \end{pmatrix}, \tag{30}$$

where the coefficients $\{E_x, E_y\}$ are real-valued numbers. The second-order current density given by Eq. (6) then reads

$$\begin{aligned} \langle \hat{j}^{(\chi)}_z(t)\rangle^{(\text{quad,Drude})} &= -\frac{e^3\chi E_x E_y}{24\pi}\beta_0(\delta\omega, \Gamma, t)F(\omega_0, \Gamma, \mu_\chi), \\ \beta_0(\delta\omega, \Gamma, t) &= \frac{\delta\omega}{\delta\omega^2 + 4\Gamma^2}\sin(\delta\omega\, t) + \frac{2\Gamma}{\delta\omega^2 + 4\Gamma^2}\cos(\delta\omega\, t), \\ F(\omega_0, \Gamma, \mu_\chi) &= \frac{1}{\pi}\text{Im}[\ln(4|\mu_\chi|^2 - (\omega_0 + i\Gamma)^2)]. \end{aligned} \tag{31}$$

In the disorder-free limit ($\Gamma \to 0^+$), we find

$$F(\omega_0, 0^+, \mu_\chi) \to -\text{sgn}(\omega_0)\Theta(|\omega_0| - 2|\mu_\chi|), \tag{32}$$

and Eq. (31) assumes the form of the quantized CPGE response discussed in Ref. [30]. We note that the limits $\Gamma \to 0^+$ and $\delta\omega \to 0^+$ do not commute, thus the quantized CPGE response is obtained by performing the limit $\Gamma \to 0^+$ first.

When $\Gamma \neq 0$, Eq. (31) indicates that the quantized CPGE response is substantially modified in the following two ways. First, the onset of a nonzero current density in terms of the optical frequency $\omega_0$ is described by the function $F(\omega_0, \Gamma, \mu_\chi)$, thus it is smeared out in a frequency range of order $\Gamma$ around $2|\mu_\chi|$. That is indicative for the existence of a *quantum rectification sum rule* associated with optical responses controlled by the Berry connection [52]. Second, the current density acquires the *Drude-like* form

$$\langle \hat{j}^{(\chi)}_z(t)\rangle^{(\text{quad,Drude})} \propto \frac{1}{\delta\omega + i2\Gamma}, \tag{33}$$

implying that, unlike in the disorder-free case, it does not grow linearly in time, but it shows an oscillatory temporal behavior determined by the function $\beta_0(\omega_0, \Gamma, \mu_\chi)$.

It is also noteworthy to point out that despite Eq. (31) having been derived under the condition $\Gamma \ll \delta\omega$, it smoothly connects to the results of Ref. [53]. In particular, for finite $\Gamma$ but $\delta\omega = 0$, the current density is time-independent, and it is proportional to the effective scattering time $\Gamma^{-1}$ for times much larger than the latter, i.e.,

$$\langle \hat{j}^{(\chi)}_z(t \gg \Gamma^{-1})\rangle^{(\text{quad,Drude})} \propto \Gamma^{-1}. \tag{34}$$

Such an agreement conforms well with the fact that Eq. (6) yields the response to an electric field that has been switched on in the infinite past, and it is hence also concerned with long-time responses. That agreement also suggests that Eq. (31) can smoothly interpolate between the limiting cases $\Gamma \gg \delta\omega$ and $\Gamma \ll \delta\omega$. Furthermore, the scaling of the current density in Eq. (34) also confirms the predicted behavior for a *rectified injection* current density when the relaxation processes are treated under a perturbative scheme [54,55].

However, it should be emphasized that the condition $\delta\omega \ll \omega_0$ must always be respected should one wish to recover the quantized CPGE result of Ref. [30] when $\Gamma = 0$. Keeping corrections of the order $\mathcal{O}(\delta\omega^3/\omega_0^3)$ in Eq. (31) for $\Gamma = 0$, Eq. (6) reads

$$\begin{aligned} \frac{\langle \hat{j}^{(\chi)}_z(t)\rangle^{(\text{quad,Drude})}}{\frac{-e^3 E_x E_y}{24\pi}} &\approx \chi\frac{\sin\delta\omega\, t}{\delta\omega}\left(1 - \frac{\delta\omega^2}{\omega_0^2}\right)F(\omega_0, \mu_\chi), \\ F(\omega_0, \mu_\chi) &= -\text{sgn}(\omega_0)\Theta(|\omega_0| - 2|\mu_\chi|). \end{aligned} \tag{35}$$

Therefore, turning away from the condition $\delta\omega \ll \omega_0$ obviously destroys the quantized CPGE response, and the temporal profile of the current density is drastically modified, even in the disorder-free case. That is so, because corrections of the order $\mathcal{O}(\delta\omega^3/\omega_0^3)$ substantially alter the resonance structure of the three-current correlation function given by Eq. (25).

## VI. NUMERICAL STUDY OF INTRANODE SCATTERING

Upon applying in Eq. (25) the prescription for analytic continuation of the bosonic Matsubara frequencies $i\omega_1 \to \omega_0 + i\eta_1$ and $i\omega_2 \to -\omega_0 + \delta\omega + i\eta_2$, one can numerically evaluate the three-current correlation function and, through Eq. (7), the two-point retarded Green's function $\mathcal{X}^{(+1,\text{SCBA})}_{\alpha\beta\gamma}(\omega_0, -\omega_o + \delta\omega)|_{\widetilde{W}_{\text{inter}}=0}$. For that purpose, we also use the approximate form of the self-energy correction given in Eq. (21) as well as the vertex correction from Eq. (18). The corresponding results for the Weyl node with chirality $\chi = +1$ are shown in Fig. 4 with the chosen values $\omega_0 = 3$ and $\delta\omega = 10^{-3}$ (from Sec. V, we recall the condition $\delta\omega \ll \omega_0$).

To compare those numerical findings to the Drude-like two-point retarded Green's function $\widetilde{\mathcal{X}}^{(+1,\text{SCBA,Drude})}_{\alpha\beta\gamma}(-\omega_0, -\omega_0 + \delta\omega)$ given by Eq. (29) with $\chi = +1$, we choose the following *ansatz* expression for the effective scattering strength $\Gamma$:

$$\Gamma = -\text{Im}\,\Sigma^{(+1)}_{\text{SCBA, approx.}}(\nu + i\eta), \tag{36}$$

where $\nu$ is an *effective* real-valued frequency, and $\eta \to 0^+$. To determine $\nu$, we analyze the momentum-frequency structure of the integrand from Eq. (25) upon neglecting vertex corrections. Then, each expression inside every pair of rectangular brackets corresponds to a spectral density function that shows global maximum at frequency

$$\nu_{\max} = b_j v_F k + \text{Re}\Sigma^{(+1)}_{\text{SCBA, approx.}}(\nu_{\max} + i\eta). \tag{37}$$

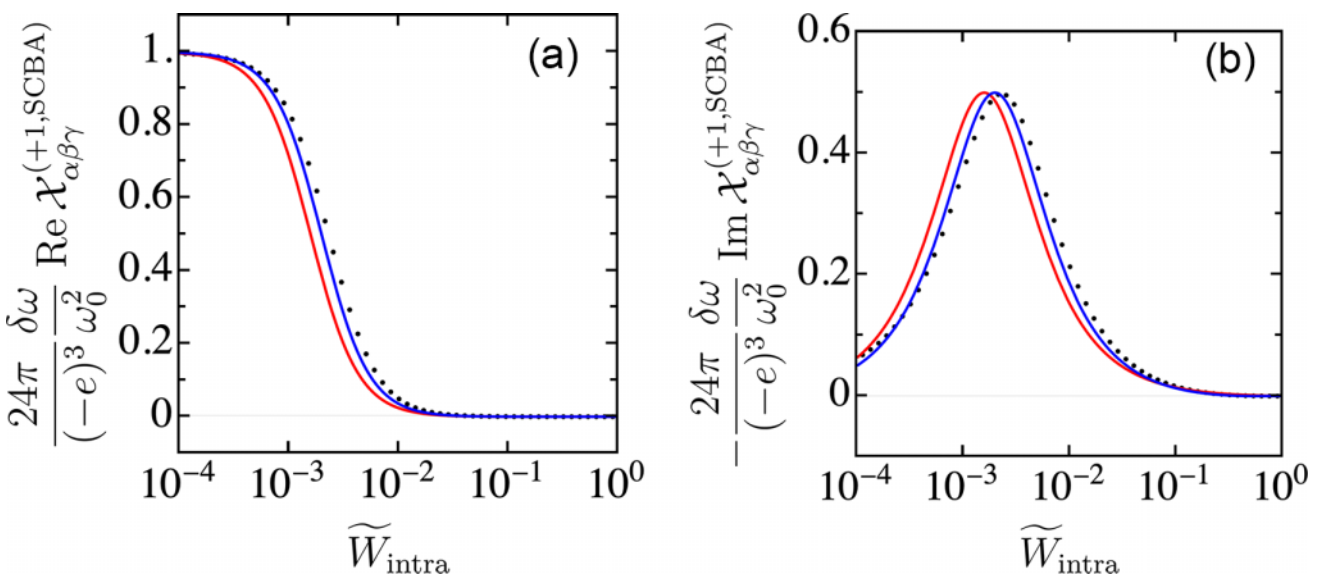

FIG. 4. (a), (b) The normalized real and imaginary parts of the two-point retarded Green's function $\mathcal{X}^{(+1,\text{SCBA})}_{\alpha\beta\gamma} \equiv \mathcal{X}^{(+1,\text{SCBA})}_{\alpha\beta\gamma}(\omega_0, -\omega_0+\delta\omega)|_{\widetilde{W}_{\text{inter}}=0}$, respectively, as functions of the dimensionless intranode scattering strength $\widetilde{W}_{\text{intra}}$ in the absence of internode scattering. The black dots show the corresponding numerical evaluation through Eqs. (7) and (25) with the replacements $i\Omega_1 \to \omega_0 + i\,\eta_1$, $i\Omega_2 \to -\omega_0+\delta\omega+i\,\eta_2$, $\omega_0 = 3$, $\delta\omega = 10^{-3}$, $\mu_1 = +1$, $v_F = 1$, $\Lambda = 25$, $\eta_1 = 1.05\times10^{-5}$, $\eta_2 = 0.95\times10^{-5}$, and $\eta = 0.9\times10^{-6}$. The red and blue lines depict the effective Drude-like two-point retarded Green's function from Eq. (29) with $\Gamma$ given by Eqs. (36) and (39), respectively. The effective frequency $\nu$ is numerically determined self-consistently via Eq. (38).

We plug such an expression for $\nu_{\max}$ into the remaining two fermionic Matsubara Green's functions outside each pair of rectangular brackets in Eq. (25), and we find that the associated integrand shows global maxima at momenta values $2v_F k_{\max} \approx \pm\omega_0$. Hence, we determine $\nu$ through the self-consistent equation

$$\nu - \text{Re}\,\Sigma^{(+1)}_{\text{SCBA, approx}}(\nu + i\eta) = \frac{\omega_0}{2}. \tag{38}$$

In physical terms, the thus evaluated Drude-like response at effective frequency $\nu$ corresponds to optical excitations of *renormalized* electronic states from an energy $-\omega_0/2$ below the Weyl node to an energy $+\omega_0/2$ above the same Weyl node. Figure 5 depicts the function $-\text{Im}\,\Sigma^{(+1)}_{\text{SCBA, approx}}(\nu+i\eta)$ in terms of the intranode scattering strength $\widetilde{W}_{\text{intra}}$ with $\nu$ determined via Eq. (38).

Regarding Fig. 4, the function $\widetilde{\mathcal{X}}^{(+1,\text{SCBA,Drude})}_{\alpha\beta\gamma}(-\omega_0, -\omega_0+\delta\omega)$, supplemented by Eq. (36), fits the numerical data quite well even close to the critical scattering regime $\widetilde{W}_{\text{intra, crit}} = 1$. The latter conclusion is another confirmation for the extension of the range of applicability of the Drude-like response given by Eq. (31) beyond the limit $\Gamma \ll \delta\omega$. As depicted in Fig. 5 for the scattering regime $\widetilde{W}_{\text{intra}} < 0.6$, the numerically obtained features of the two-point retarded Green's function $\mathcal{X}^{(+1,\text{SCBA})}_{\alpha\beta\gamma}(\omega_0, -\omega_o+\delta\omega)|_{\widetilde{W}_{\text{inter}}=0}$ can also be described (in fact even slightly better) by the further approximate formulas [51]:

$$\text{Re}\,\Sigma^{(\chi)}_{\text{SCBA, approx}}(\omega) = -\frac{\widetilde{W}_{\text{intra}}\,\omega}{1-\widetilde{W}_{\text{intra}}},$$
$$\Gamma = \frac{\pi\,\widetilde{W}_{\text{intra}}\,\omega^2}{2\Lambda(1-\widetilde{W}_{\text{intra}})^3}. \tag{39}$$

In conjuction with Eq. (38), the expressions in Eq. (39) can be used to estimate two quantities. First, the global maximum of the imaginary part of $\mathcal{X}^{(+1,\text{SCBA})}_{\alpha\beta\gamma}(\omega_0, -\omega_o+\delta\omega)|_{\widetilde{W}_{\text{inter}}=0}$

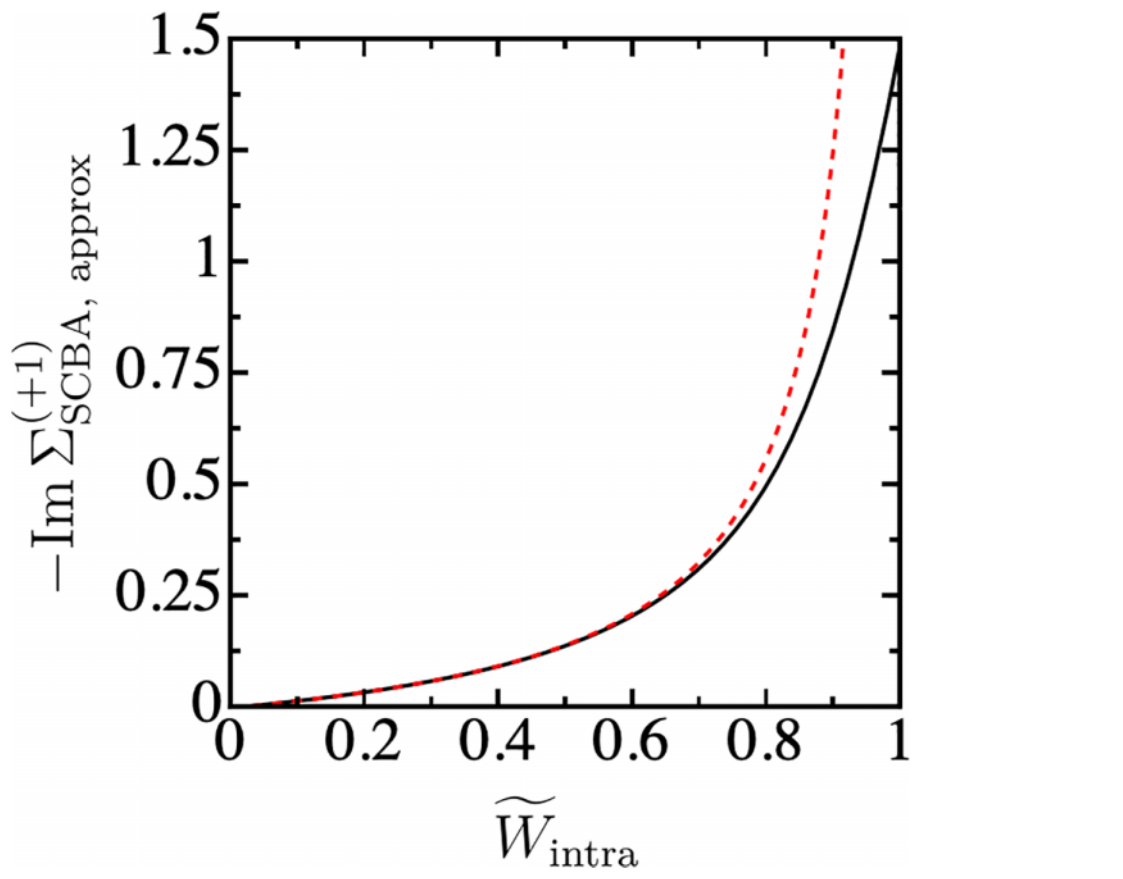

FIG. 5. The black solid line shows the part $-\text{Im}\,\Sigma^{(\chi)}_{\text{SCBA, approx.}}(\nu+i\eta)$ of the self-energy correction given by Eq. (21) as a function of the dimensionless scattering strength $\widetilde{W}_{\text{intra}}$. The real frequency $\nu$ has been evaluated self-consistently via Eq. (38). The red dashed line shows the estimation for $-\text{Im}\,\Sigma^{(\chi)}_{\text{SCBA, approx.}}(\nu+i\eta)$ when the approximate formulas from Eq. (39) are used for the self-energy correction. The numerical parameters used are $v_F = 1$, $\Lambda = 25$, and $\omega_0 = 3$.

shown in Figs. 4 and 6 is analytically yet approximately estimated to be

$$-\frac{24\pi}{(-e)^3}\frac{\delta\omega}{\omega_0^2}\text{Im}\,\mathcal{X}^{(+1,\text{SCBA})}_{\alpha\beta\gamma,\text{cross}} = \frac{\frac{\pi\,\delta\omega\widetilde{W}_{\text{intra,cross}}\omega_0^2}{4\Lambda(1-\widetilde{W}_{\text{intra,cross}})^3}}{(\delta\omega)^2 + \left[\frac{\pi\widetilde{W}_{\text{intra,cross}}\omega_0^2}{4\Lambda(1-\widetilde{W}_{\text{intra,cross}})^3}\right]^2}. \tag{40}$$

Second, that global maximum is predicted to occur at the scattering strength

$$\widetilde{W}_{\text{intra,cross}} = \frac{1}{1+\frac{\pi\omega_0^2}{4\delta\omega\Lambda}}. \tag{41}$$

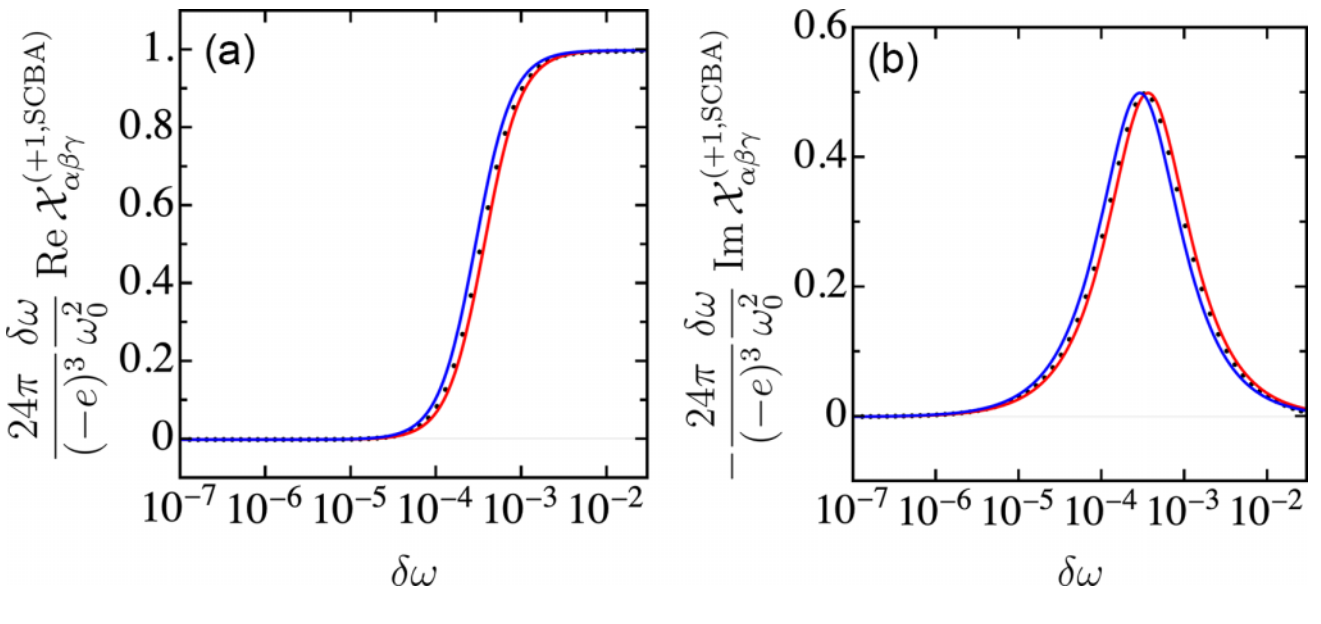

FIG. 6. (a), (b) The normalized real and imaginary parts of the two-point retarded Green's function $\mathcal{X}^{(+1,\text{SCBA})}_{\alpha\beta\gamma} \equiv \mathcal{X}^{(+1,\text{SCBA})}_{\alpha\beta\gamma}(\omega_0, -\omega_0+\delta\omega)|_{\widetilde{W}_{\text{inter}}=0}$, respectively, as functions of the frequency detuning $\delta\omega$ in the absence of internode scattering. The black dots show the corresponding numerical evaluation through Eqs. (7) and (25) with the replacements $i\Omega_1 \to \omega_0 + i\,\eta_1$, $i\Omega_2 \to -\omega_0+\delta\omega+i\,\eta_2$, $\omega_0 = 3$, $\widetilde{W}_{\text{intra}} = 10^{-3}$, $\mu_1 = +1$, $v_F = 1$, $\Lambda = 25$, $\eta_1 = 1.05\times10^{-5}$, $\eta_2 = 0.95\times10^{-5}$, and $\eta = 0.9\times10^{-6}$. The red and blue lines depict the effective Drude-like two-point retarded Green's function from Eq. (29) with $\Gamma$ given by Eqs. (36) and (39), respectively. The effective frequency $\nu$ is numerically determined self-consistently via Eq. (38).

Determining the values $\mathrm{Im}\,\mathcal{X}^{(+1,\mathrm{SCBA})}_{\alpha\beta\gamma,\mathrm{cross}}$ and $\widetilde{W}_{\mathrm{intra,cross}}$ (even approximately) is significant, because they define a *crossover scale* between the response regimes $\delta\omega > 2\Gamma$ and $\delta\omega < 2\Gamma$. In the former regime, the imaginary part of $\mathcal{X}^{(+1,\mathrm{SCBA})}_{\alpha\beta\gamma}(\omega_0, -\omega_o + \delta\omega)|_{\widetilde{W}_{\mathrm{inter}}=0}$ grows monotonically as a function of intranode scattering strength, and it smoothly connects to the limiting case $\delta\omega \gg \Gamma$ characterized by the current-density profile

$$\left\langle j_z^{(+1)}(t)\right\rangle^{(\mathrm{quad,Drude})} \propto \delta\omega^{-1}\,\sin(\delta\omega\, t). \tag{42}$$

On the other hand, when $\delta\omega < 2\Gamma$, decay sets in for the aforementioned imaginary part, and the corresponding limiting case $\delta\omega \ll \Gamma$ is predicted to have the current-density profile

$$\left\langle j_z^{(+1)}(t)\right\rangle^{(\mathrm{quad,Drude})} \propto \Gamma^{-1}\,\cos(\delta\omega\, t). \tag{43}$$

A complementary viewpoint to Fig. 4 is the case where the two-point retarded Green's function $\mathcal{X}^{(+1,\mathrm{SCBA})}_{\alpha\beta\gamma}(\omega_0, -\omega_o + \delta\omega)|_{\widetilde{W}_{\mathrm{inter}}=0}$ is numerically evaluated for fixed intranode scattering $\widetilde{W}_{\mathrm{intra}}$ but for varied frequency detuning $\delta\omega$. That scenario is depicted in Fig. 6. As in Fig. 4, we observe a similarly good agreement between the Drude-like expression given by Eq. (29) and the numerically evaluated $\mathcal{X}^{(+1,\mathrm{SCBA})}_{\alpha\beta\gamma}(\omega_0, -\omega_o + \delta\omega)|_{\widetilde{W}_{\mathrm{inter}}=0}$, including also the regime far beyond the limit $\Gamma \ll \delta\omega$. Furthermore, we note that the amplitude of the part $\mathrm{Re}\,\mathcal{X}^{(+1,\mathrm{SCBA})}_{\alpha\beta\gamma}(\omega_0, -\omega_o + \delta\omega)|_{\widetilde{W}_{\mathrm{inter}}=0}$ reaches a "quantized CPGE"-like value in the regime $\delta\omega \gg \Gamma$, provided the condition $\delta\omega \ll \omega_0$ is satisfied. However, as far as the current density in Eq. (6) is concerned, its temporal profile is an oscillatory function with the oscillation period determined by the frequency detuning $\delta\omega$, as the Drude-like model in Eq. (31) suggests. Therefore, such a "quantized CPGE"-like response is in sharp contrast to the quantized CPGE response in the disorder-free limit, which is associated with a current density growing linearly in time [30].

Thus, considering Figs. 4 and 6, we conclude that the entire second-order response in the presence of intranode scattering only can be determined through the $\pi$ shift and the distinct dependence of the current density on the frequency detuning $\delta\omega$ in the two regimes $\delta\omega \ll \Gamma$ and $\delta\omega \gg \Gamma$. That conclusion should provide clear experimental signatures of the predicted Drude-like behavior. More generally, the amplitude and the phase (with respect to the elliptically polarized external electric field) of the current density as a function of $\delta\omega$ and $\Gamma$ should also enable a complete experimental determination of the two-point retarded Green's function $\mathcal{X}^{(+1,\mathrm{SCBA})}_{\alpha\beta\gamma}(\omega_0, -\omega_o + \delta\omega)|_{\widetilde{W}_{\mathrm{inter}}=0}$.

## VII. EFFECT OF INTERNODE SCATTERING

To estimate the effect of internode scattering to the current density given by Eq. (6), we numerically evaluate the two-point retarded Green's function $\mathcal{X}^{(+1,\mathrm{SCBA})}_{\alpha\beta\gamma}(\omega_0, -\omega_o + \delta\omega)$ similarly to Sec. VI but with $\widetilde{W}_{\mathrm{inter}} \neq 0$. We enforce the Pauli-blocking condition at the Weyl node with chirality $\chi = -1$ by assuming that the corresponding energy offset $\mu_{-1}$ is sufficiently larger compared to $\omega_0/2$, since impurity scattering tends to smear out (and thus modify) the response function around the onset frequency $2|\mu_{-1}|$. When a linear interdependence with fixed proportionality factor is assumed between the strengths of inter- and intranode scattering, i.e., $\widetilde{W}_{\mathrm{inter}} \propto \widetilde{W}_{\mathrm{intra}}$, the combined effect of intra- and internode scattering on the function $\mathcal{X}^{(+1,\mathrm{SCBA})}_{\alpha\beta\gamma}(\omega_0, -\omega_o + \delta\omega)$ can be controlled by tuning the strength of intranode scattering $\widetilde{W}_{\mathrm{intra}}$ only. That is shown in Fig. 7 for different proportionality factors. When such linear interdependence is present, the value of $\widetilde{W}_{\mathrm{intra}}$ can be viewed as a proxy for an effective scattering strength comprising the effects of $\widetilde{W}_{\mathrm{intra}}$ and $\widetilde{W}_{\mathrm{inter}}$. Thus, such a scenario corresponds to an effective intranode scattering case with a rescaled scattering strength. Consequently, the numerical data in Fig. 7 qualitatively preserve the features of the Drude-like response even for strong internode scattering. However, in the latte regime, the part $\mathrm{Im}\,\mathcal{X}^{(+1,\mathrm{SCBA})}_{\alpha\beta\gamma}(\omega_0, -\omega_o + \delta\omega)$ shows suppressed peak values and broadened bandwidths, while the entire response has been shifted toward smaller intranode scattering strengths. Another interesting feature of Fig. 7 is the nonequidistant spacing between data points of different proportionality factors indicating a nonlinear scaling of $\mathcal{X}^{(+1,\mathrm{SCBA})}_{\alpha\beta\gamma}(\omega_0, -\omega_o + \delta\omega)$ as a function of the internode scattering strength.

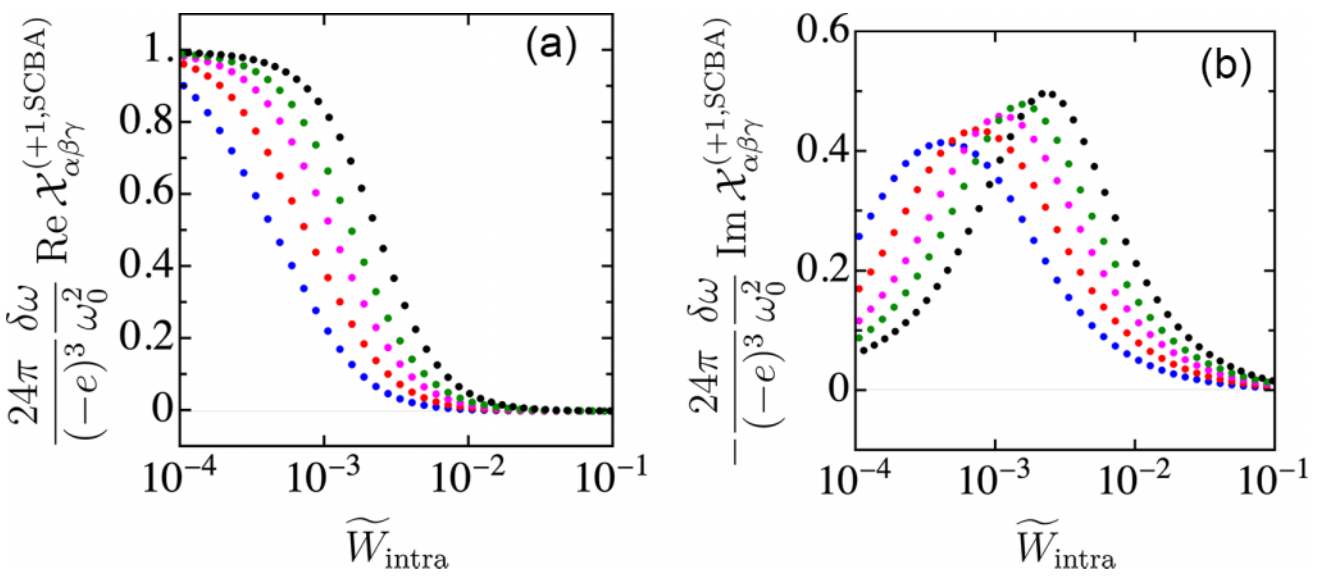

FIG. 7. (a), (b) The normalized real and imaginary parts of the two-point retarded Green's function $\mathcal{X}^{(+1,\mathrm{SCBA})}_{\alpha\beta\gamma} \equiv \mathcal{X}^{(+1,\mathrm{SCBA})}_{\alpha\beta\gamma}(\omega_0, -\omega_0 + \delta\omega)$, respectively, as functions of the dimensionless intranode scattering strength $\widetilde{W}_{\mathrm{intra}}$ for four different scenarios for the ratio $x = \widetilde{W}_{\mathrm{inter}}/\widetilde{W}_{\mathrm{intra}}$ when $\mu_{-1} = 3$: $x = 1$ (blue dots), $x = 0.5$ (red dots), $x = 0.25$ (magenta dots), and $x = 0.125$ (green dots). The black dots are the numerical data from Fig. 4 in the absence of internode scattering. The numerical evaluation has been done using the same values as Fig. 4 for the remaining parameters.

For the particular linear interdependence $\widetilde{W}_{\mathrm{inter}} = 0.125 \widetilde{W}_{\mathrm{intra}}$ and for varied energy offset $\mu_{-1}$ of the photo-idle Weyl node, Fig. 8 shows the function $\mathcal{X}^{(+1,\mathrm{SCBA})}_{\alpha\beta\gamma}(\omega_0, -\omega_o + \delta\omega)$ to still preserve the features of the Drude-like response, though the entire complex-valued function is shifted toward smaller intranode scattering strengths as the energy difference between the two Weyl nodes grows, with the part $\mathrm{Im}\,\mathcal{X}^{(+1,\mathrm{SCBA})}_{\alpha\beta\gamma}(\omega_0, -\omega_o + \delta\omega)$ to be slightly distorted. We also note the strong nonlinear dependence of the function $\mathcal{X}^{(+1,\mathrm{SCBA})}_{\alpha\beta\gamma}(\omega_0, -\omega_o + \delta\omega)$ on the energy difference between the Weyl nodes. On the oher hand, Fig. 9 illustrates that both parts of the two-point retarded Green's function are reaching nonzero plateau values depending on the strength of internode scattering, when $\widetilde{W}_{\mathrm{intra}} < \widetilde{W}_{\mathrm{inter}}$. As far as the part $\mathrm{Re}\,\mathcal{X}^{(+1,\mathrm{SCBA})}_{\alpha\beta\gamma}(\omega_0, -\omega_o + \delta\omega)$ is concerned, its response and plateau value depend more strongly on the strength of internode scattering compared to $\mathrm{Im}\,\mathcal{X}^{(+1,\mathrm{SCBA})}_{\alpha\beta\gamma}(\omega_0, -\omega_o + \delta\omega)$. Consequently, such a scattering regime *cannot* be described in terms of our simplistic Drude-like model given by Eq. (31). Moreover, Fig. 10 shows that nonzero plateau values for $\mathcal{X}^{(+1,\mathrm{SCBA})}_{\alpha\beta\gamma}(\omega_0, -\omega_o + \delta\omega)$ can be also reached at small values of $\widetilde{W}_{\mathrm{intra}}$ for varied $\mu_{-1}$ at fixed $\widetilde{W}_{\mathrm{inter}}$. In that case, we note that the suppression of the two-point retarded Green's function across the entire scattering regime is stronger compared to Fig. 9.

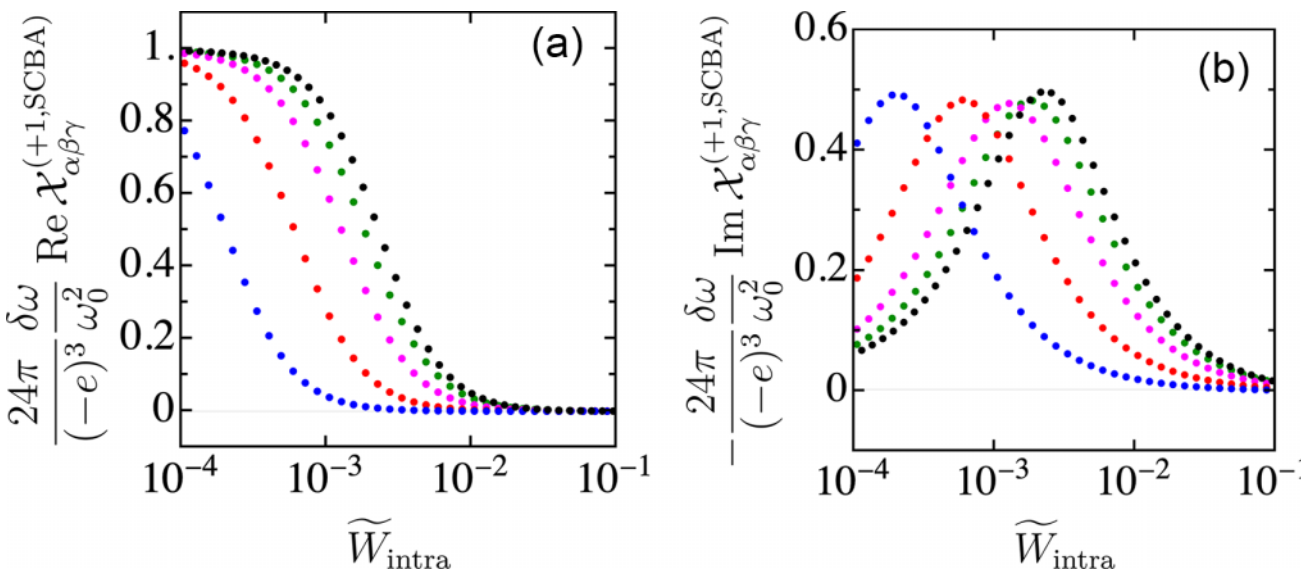

FIG. 8. (a), (b) The normalized real and imaginary parts of the two-point retarded Green's function $\mathcal{X}^{(+1,\mathrm{SCBA})}_{\alpha\beta\gamma} \equiv \mathcal{X}^{(+1,\mathrm{SCBA})}_{\alpha\beta\gamma}(\omega_0, -\omega_0 + \delta\omega)$, respectively, as functions of the dimensionless intranode scattering strength $\widetilde{W}_{\mathrm{intra}}$ when $\widetilde{W}_{\mathrm{inter}} = 0.125\widetilde{W}_{\mathrm{intra}}$ for four different scenarios of the energy offset of the second Weyl node from the chemical potential: $\mu_{-1} = 16$ (blue dots), $\mu_{-1} = 8$ (red dots), $\mu_{-1} = 4$ (magenta dots), and $\mu_{-1} = 2$ (green dots). The black dots are the numerical data from Fig. 4 in the absence of internode scattering. The numerical evaluation has been done using the same values as Fig. 4 for the remaining parameters.

Finally, we present numerical data in Figs. 11 and 12 indicating that the increase of the frequency detuning $\delta\omega$ tends to recover the quantized value for $\mathrm{Re}\,\mathcal{X}^{(+1,\mathrm{SCBA})}_{\alpha\beta\gamma}(\omega_0, -\omega_o + \delta\omega)$, albeit the corresponding current density oscillates periodically in time due to the finite frequency detuning. Similarly to the discussion in Sec. VI, we expect the process of fine-tuning $\delta\omega$ to depend on the optical frequency $\omega_0$, as well as on other relevant system parameters, such as the intra- and internode scattering strengths, and the energy difference between the Weyl nodes.

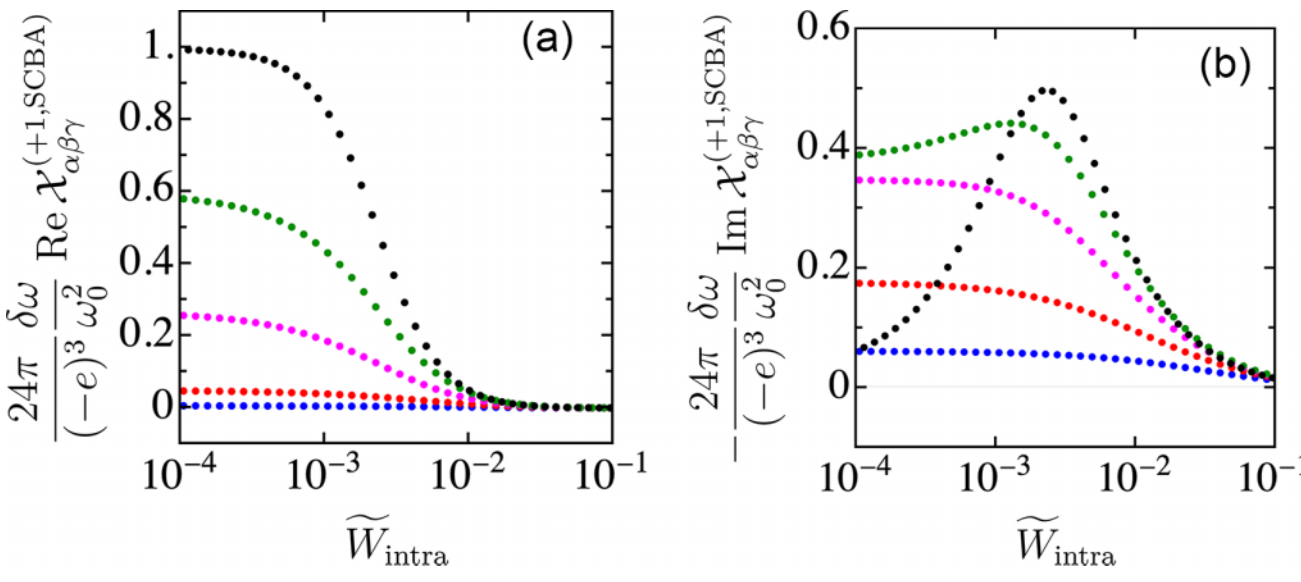

FIG. 9. (a), (b) The normalized real and imaginary parts of the two-point retarded Green's function $\mathcal{X}^{(+1,\mathrm{SCBA})}_{\alpha\beta\gamma} \equiv \mathcal{X}^{(+1,\mathrm{SCBA})}_{\alpha\beta\gamma}(\omega_0, -\omega_0 + \delta\omega)$, respectively, as functions of the dimensionless intranode scattering strength $\widetilde{W}_{\mathrm{intra}}$ for four different scenarios of internode scattering strength $\widetilde{W}_{\mathrm{inter}}$ when $\mu_{-1} = 3$: $\widetilde{W}_{\mathrm{inter}} = 10^{-2}$ (blue dots), $\widetilde{W}_{\mathrm{inter}} = 10^{-2.5}$ (red dots), $\widetilde{W}_{\mathrm{inter}} = 10^{-3}$ (magenta dots), and $\widetilde{W}_{\mathrm{inter}} = 10^{-3.5}$ (green dots). The black dots are the numerical data from Fig. 4 in the absence of internode scattering. The numerical evaluation has been done using the same values as Fig. 4 for the remaining parameters.

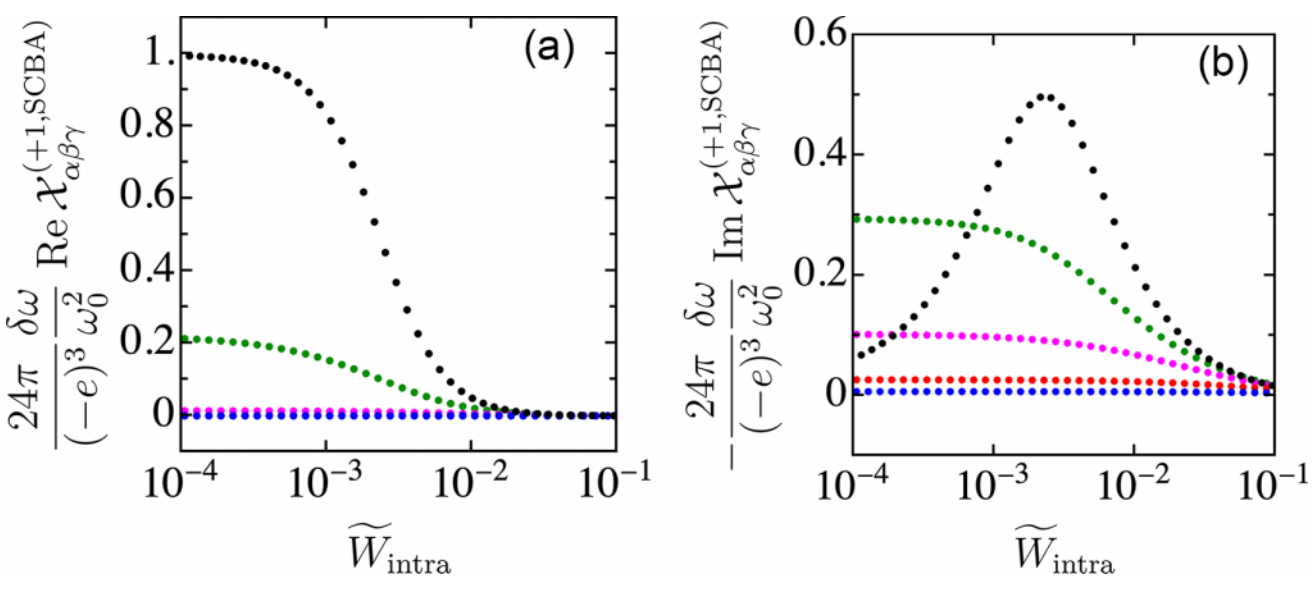

FIG. 10. (a), (b) The normalized real and imaginary parts of the two-point retarded Green's function $\mathcal{X}^{(+1,\mathrm{SCBA})}_{\alpha\beta\gamma} \equiv \mathcal{X}^{(+1,\mathrm{SCBA})}_{\alpha\beta\gamma}(\omega_0, -\omega_0 + \delta\omega)$, respectively, as functions of the dimensionless intranode scattering strength $\widetilde{W}_{\mathrm{intra}}$ for four different scenarios for the energy offset of the photo-idle Weyl node from the chemical potential when $\widetilde{W}_{\mathrm{inter}} = 10^{-3}$: $\mu_{-1} = 16$ (blue dots), $\mu_{-1} = 8$ (red dots), $\mu_{-1} = 4$ (magenta dots), and $\mu_{-1} = 2$ (green dots). The black dots are the numerical data from Fig. 4 in the absence of internode scattering. The numerical evaluation has been done using the same values as Fig. 4 for the remaining parameters.

## VIII. SUMMARY AND OUTLOOK

In summary, we have generalized the second-order optical response of the CPGE in the presence of finite frequency detuning on a system with two Weyl nodes located at different energies. We considered the behavior of the real and imaginary parts of the two-point retarded Green's function in terms of pointlike elastic impurity intra- and internode scattering, frequency detuning, and energy difference between the two Weyl nodes. We have presented numerical data in Figs. 4 and 5 substantiated by the effective model for the current density given in Eq. (31) suggesting that the entire second-order response as a function of intranode scattering is qualitatively similar to the Drude-like behavior of the AC linear optical conductivity in metals. As long as intra- and internode scattering strengths are linearly interdependent, the Drude-like characteristics of the optical response are preserved, as demonstrated in Figs. 7 and 8. However, when the internode scattering strength is kept fixed while the intranode scattering strength varies, the current-density profile deviates strongly from the Drude-like behavior, and it shows nonzero plateau values for both the real and imaginary parts of the corresponding two-point retarded Green's function in the limit of weak intranode scattering (see Figs. 9 and 10). We have also presented data in Figs. 6, 11, and 12 suggesting that the real part of the two-point retarded Green's function can reach a "quantized" value when the frequency detuning is properly adjusted.

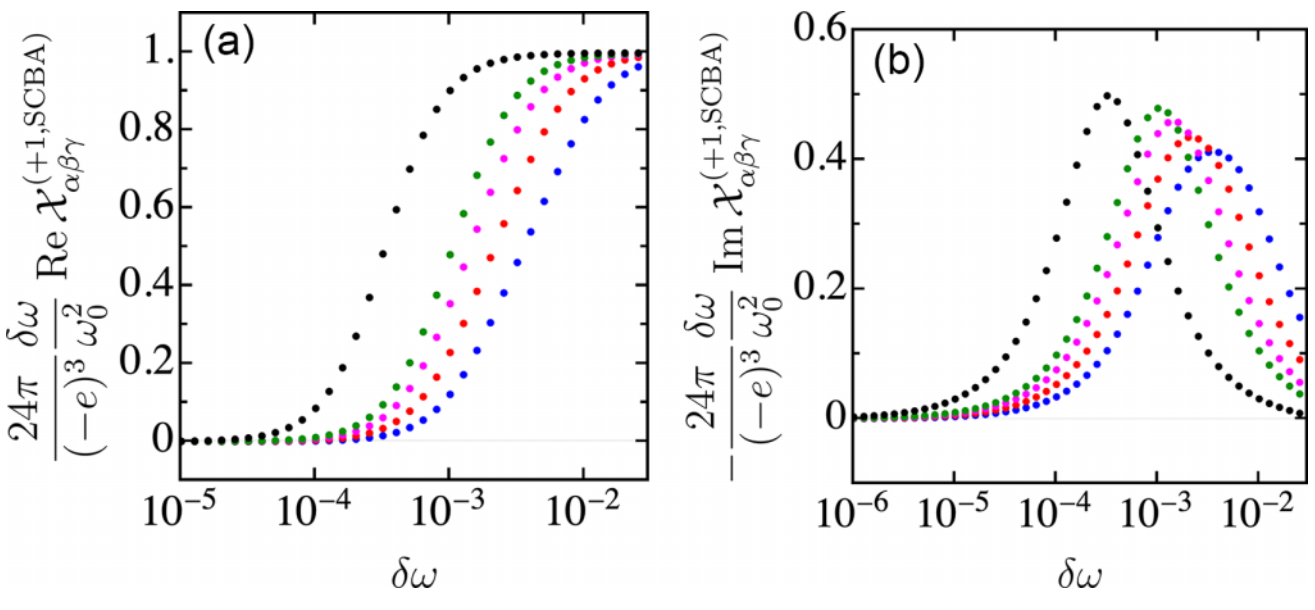

FIG. 11. (a), (b) The normalized real and imaginary parts of the two-point retarded Green's function $\mathcal{X}^{(+1,\mathrm{SCBA})}_{\alpha\beta\gamma} \equiv \mathcal{X}^{(+1,\mathrm{SCBA})}_{\alpha\beta\gamma}(\omega_0, -\omega_0 + \delta\omega)$, respectively, as functions of the frequency detuning $\delta\omega$ for four different scenarios for the ratio $x = \widetilde{W}_{\mathrm{inter}}/\widetilde{W}_{\mathrm{intra}}$ when $\mu_{-1} = 3$: $x = 1$ (blue dots), $x = 0.5$ (red dots), $x = 0.25$ (magenta dots), and $x = 0.125$ (green dots). The black dots are the numerical data from Fig. 4 in the absence of internode scattering. The numerical evaluation has been done using the same values as Fig. 4 for the remaining parameters.

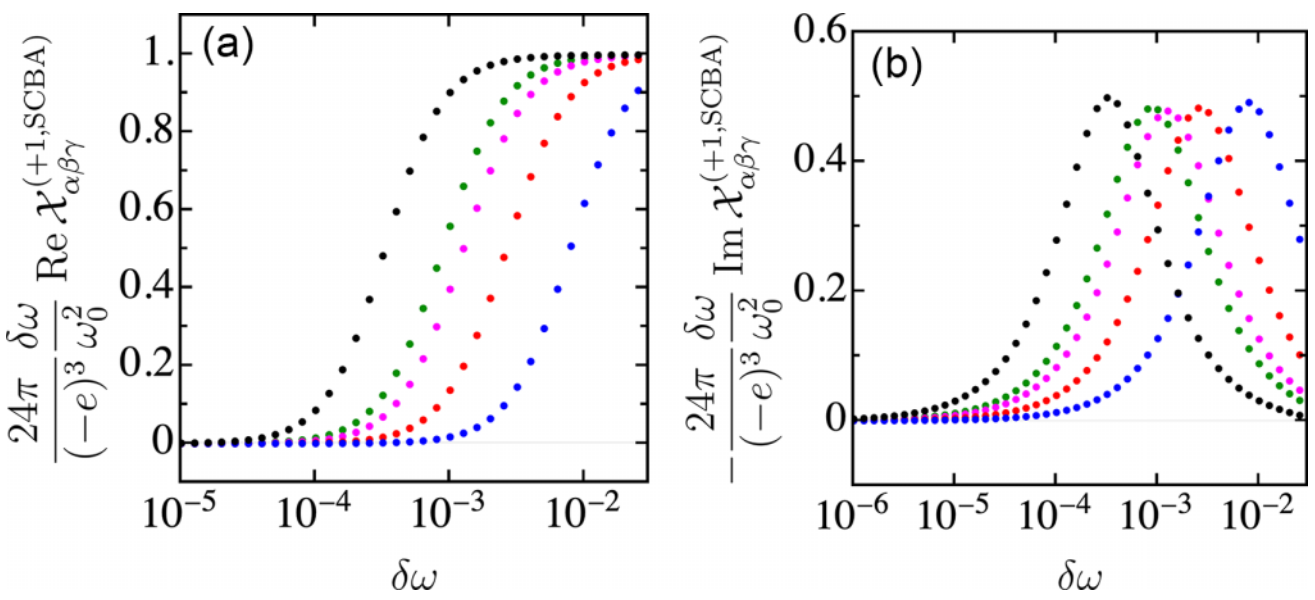

FIG. 12. (a), (b) The normalized real and imaginary parts of the two-point retarded Green's function $\mathcal{X}^{(+1,\mathrm{SCBA})}_{\alpha\beta\gamma} \equiv \mathcal{X}^{(+1,\mathrm{SCBA})}_{\alpha\beta\gamma}(\omega_0, -\omega_0 + \delta\omega)$, respectively, as functions of the frequency detuning $\delta\omega$ for four different scenarios for the energy offset of the second Weyl node from the chemical potential when $\widetilde{W}_{\mathrm{inter}}/\widetilde{W}_{\mathrm{intra}} = 0.125$: $\mu_{-1} = 16$ (blue dots), $\mu_{-1} = 8$ (red dots), $\mu_{-1} = 4$ (magenta dots), and $\mu_{-1} = 2$ (green dots). The black dots are the numerical data from Fig. 4 in the absence of internode scattering. The numerical evaluation has been done using the same values as Fig. 4 for the remaining parameters.

We conclude the present paper with an outlook on the experimental relevance of the aforementioned findings. The distinctive consequences of the finite frequency detuning and impurity scattering on the second-order current density given by Eq. (6) make the case that a nontrivial dependence can be probed by all externally controlled parameters, namely, the density of impurity scattering centers, the frequency detuning, the optical frequency, and the energy difference between the Weyl nodes. Moreover, measuring the amplitude and phase shift of the current density with respect to the applied electric field, one can determine the real and imaginary parts of the two-point retarded Green's function, and hence, by comparing to our results in Secs. VI and VII, the effective scattering time $\Gamma^{-1}$—at least in a regime where the Drude-like model is valid. Candidate materials where those experimental probes can be implemented are transition metal monopnictides (TMMPs) such as TaAs, TaP, and NbAs. TMMPs have already been used to identify the existence of CPGE in Weyl semimetals through the measurement of the helicity-dependent generated voltage [39], to probe the giant nonlinear response associated with the second harmonic generation [56], and to detect the chirality of Weyl fermions through the CPGE response using mid-infrared optical frequency [38]. Furthermore, the predictions discussed in Secs. VI and VII can also be tested in the semimetal CoSi, where the quantized CPGE response has been already hinted to appear, as well as in other materials, such as heterostructures [57], alloys [58], and half-Heusler compounds [59]. Finally, we expect our findings to be applicable also to materials with multiple pairs of Weyl nodes, provided they are well separated in momentum domain.

## ACKNOWLEDGMENTS

The authors are grateful for stimulating discussions with Steffen Sykora in the initial stages of this work. They acknowledge funding by the Deutsche Forschungsgemeinschaft (DFG) via the Emmy Noether Programme (Quantum Design Grant No. ME4844/1, Project No. 327807255), Project A04 of the Collaborative Research Center SFB 1143 (Project No. 247310070), and the Cluster of Excellence on Complexity and Topology in Quantum Matter ct.qmat (EXC 2147, Project No. 390858490). The authors are furthermore grateful to the Center for Information Services and High Performance Computing (ZIH) at TU Dresden for providing computing time.

## DATA AVAILABILITY

No data were created or analyzed in this study.

## APPENDIX A: SUMMARY OF QUADRATIC RESPONSE THEORY

Here, we provide a short review on the derivation of the second-order current density. To begin with, one performs a series expansion on the $\alpha$th component of the current-density operator $\hat{j}_\alpha$ in terms of the applied electromagnetic potential [24,25]. The Greek index $\alpha$ runs through the spatial coordinates $\{x, y, z\}$. Upon implementing the *velocity gauge* [60], the series expansion yields $\hat{j}_\alpha = \sum_n \hat{j}^{(n)}_\alpha$, with the expression

$$\hat{j}^{(n)}_\alpha = \frac{e}{\hbar n!}\widehat{D}_\alpha \prod_{i=1}^{n} \frac{eA_{\alpha_i}\widehat{D}_{\alpha_i}}{\hbar} \tag{A1}$$

defines the $n$th-order current-density operator in terms of the electromagnetic vector potential $A_{\alpha_i}$ and covariant derivative $\widehat{D}_{\alpha_i}$, where the Greek indices $\alpha_{\{1,\ldots,n\}}$ run through the spatial coordinates $\{x, y, z\}$. While the diamagnetic term in the current-density operator of a free electron gas is a well-known example for the "linear term" $\hat{j}^{(1)}_\alpha$ [48], lattice-based Hamiltonian operators generically feature contributions at all orders of $n$. Suppressing integrations, prefactors, space-time indicators, and summation indices, the *second-order current density* can schematically be expressed as

$$\begin{aligned}\langle \hat{j}_\alpha(\boldsymbol{r}, t)\rangle^{(\mathrm{quad})} = {}& \langle \hat{j}^{(2)}_\alpha\rangle_0 + \big\langle\big[\hat{j}^{(1)}_\alpha, \hat{H}^{(1)}_{\mathrm{pert}}\big]\big\rangle_0 + \big\langle\big[\hat{j}^{(0)}_\alpha, \hat{H}^{(2)}_{\mathrm{pert}}\big]\big\rangle_0 \\ & + \big\langle\big[\hat{j}^{(0)}_\alpha, \big[\hat{H}^{(1)}_{\mathrm{pert}}, \hat{H}^{(1)}_{\mathrm{pert}}\big]\big]\big\rangle_0.\end{aligned} \tag{A2}$$

The symbol $\langle\cdot\rangle_0$ denotes an expectation value with respect to the time-independent Hamiltonian operator $\hat{H}_0$ in the absence of any electromagnetic vector potential, while the operator $\hat{H}^{(n)}_{\mathrm{pert}} = -\hat{j}^{(n-1)}_\alpha \cdot \hat{A}_\alpha$ is a vertex term describing the coupling of the time-independent system to the electromagnetic vector potential. The second-order term $\langle \hat{j}_\alpha^{(2)} \rangle_0$ is called *Drude-weight dipole*, and the rest of terms in Eq. (A2) correspond to one- and two-photon resonances. In particular, the CPGE is a part of the current-density operator $\langle [\hat{j}_\alpha^{(0)}, [\hat{H}_{\text{pert}}^{(1)}, \hat{H}_{\text{pert}}^{(1)}]] \rangle_0$ [60].

Similarly to the linear response theory, the second-order response can also be understood as describing the temporal dynamics of equilibrium systems in the presence of a *weak* and time-dependent electromagnetic vector potential long after the perturbation has been switched on. Therefore, one may connect the two-point retarded Green's function underlying the second-order response to the three-current correlation function, which is an equilibrium Matsubara Green's function. Restoring all notational details and adopting the velocity gauge after approximating the model to its low-energy form, the expectation value of the second-order current-density operator can be written as [61]

$$\langle \hat{j}_\alpha(\boldsymbol{r}, t) \rangle^{(\text{quad})} = \int_{-\infty}^{\infty} dt' \int d^3r' \int_{-\infty}^{\infty} dt \int d^3r'' \sum_{\beta,\gamma} X_{\alpha\beta\gamma}(\boldsymbol{r}-\boldsymbol{r}', \boldsymbol{r}-\boldsymbol{r}''; t-t', t-t'') A_\beta(\boldsymbol{r}', t') A_\gamma(\boldsymbol{r}'', t''). \tag{A3}$$

In the latter expression, the Greek indices $\{\alpha, \beta, \gamma\}$ run through the spatial coordinates, and the electromagnetic vector potential has been treated classically. Furthermore, the two-point retarded Green's function is defined as

$$X_{\alpha\beta\gamma}(\boldsymbol{r}-\boldsymbol{r}', \boldsymbol{r}-\boldsymbol{r}''; t-t', t-t'') := -\Theta(t-t')\Theta(t'-t'') \langle [[\hat{j}_\alpha(\boldsymbol{r}, t), \hat{j}_\beta(\boldsymbol{r}', t')], \hat{j}_\gamma(\boldsymbol{r}'', t'')] \rangle_0. \tag{A4}$$

Upon symmetrizing the function $X_{\alpha\beta\gamma}(\boldsymbol{r}-\boldsymbol{r}', \boldsymbol{r}-\boldsymbol{r}''; t-t', t-t'')$ through the relabelling of indices

$$\beta \leftrightarrow \gamma, \quad t' \leftrightarrow t'', \quad \boldsymbol{r}' \leftrightarrow \boldsymbol{r}'', \tag{A5}$$

the Fourier transform of the *symmetrized* two-point retarded Green's function reads

$$\mathcal{X}_{\alpha\beta\gamma}(\boldsymbol{q}_1, \boldsymbol{q}_2; \omega_1, \omega_2) = \int d^3r_1 \int_{-\infty}^{+\infty} dt_1 \int d^3r_2 \int_{-\infty}^{+\infty} dt_2\, e^{-i(\boldsymbol{q}_1\cdot\boldsymbol{r}_1 - \omega_1 t_1 + \boldsymbol{q}_2\cdot\boldsymbol{r}_2 - \omega_2 t_2)} \frac{X_{\alpha\beta\gamma}(\boldsymbol{r}_1, \boldsymbol{r}_2; t_1, t_2) + X_{\alpha\gamma\beta}(\boldsymbol{r}_2, \boldsymbol{r}_1; t_2, t_1)}{2}, \tag{A6}$$

where spatio-temporal translational invariance has been assumed. Via Lehmann representation [49], Eq. (A6) can be connected to the three-current correlation function $\widetilde{\mathcal{X}}_{\alpha\beta\gamma}(\boldsymbol{q}_1, \boldsymbol{q}_2; i\Omega_1, i\Omega_2)$:

$$\mathcal{X}_{\alpha\beta\gamma}(\boldsymbol{q}_1, \boldsymbol{q}_2; \omega_1, \omega_2) = \tfrac{1}{2}\, \widetilde{\mathcal{X}}_{\alpha\beta\gamma}(\boldsymbol{q}_1, \boldsymbol{q}_2; i\Omega_1, i\Omega_2)|_{i\Omega_{1,2} \to \omega_{1,2} + i\eta_{1,2}}. \tag{A7}$$

In the latter expression, parameters $\Omega_{1,2}$ are bosonic Matsubara frequencies, and $\eta_{1,2}$ are infinitesimal factors. The three-current correlation function is given by

$$\widetilde{\mathcal{X}}_{\alpha\beta\gamma}(\boldsymbol{q}_1, \boldsymbol{q}_2; i\Omega_1, i\Omega_2) = \int d^3r_1 \int_0^\beta d\tau_1 \int d^3r_2 \int_0^\beta d\tau_2\, e^{-i(\boldsymbol{q}_1\cdot\boldsymbol{r}_1 - \Omega_1\tau_1 + \boldsymbol{q}_2\cdot\boldsymbol{r}_2 - \Omega_2\tau_2)} \langle T_\tau\, \hat{J}_\alpha(0,0) \hat{J}_\beta(-\boldsymbol{r}_1, -\tau_1) \hat{J}_\gamma(-\boldsymbol{r}_2, -\tau_2) \rangle. \tag{A8}$$

Parameter $\tau$ represents the imaginary time, and the quantities $\hat{J}_{\{\alpha,\beta,\gamma\}}$ are the current-density operators on the imaginary-time formalism. Within the Heisenberg picture, one finds that $\hat{J}_{\{\alpha,\beta,\gamma\}}(\boldsymbol{r}, \tau=0) = \hat{j}_{\{\alpha,\beta,\gamma\}}(\boldsymbol{r}, t=0) = \hat{j}_{\{\alpha,\beta,\gamma\}}(\boldsymbol{r})$. Expressing the electromagnetic vector potential in Eq. (A3) in terms of an electric field, and assuming the latter to be *elliptically polarized* and spatially homogeneous, i.e., $\boldsymbol{E}(t) = \sum_{n=1}^2 \boldsymbol{E}_n e^{-i\omega_n t}$ with real-valued coefficients $|\boldsymbol{E}_1| \neq |\boldsymbol{E}_2|$, the expectation value of the second-order current-density operator then reads

$$\langle \hat{j}_\alpha(\boldsymbol{r}, t) \rangle^{(\text{quad})} = \sum_{\beta,\gamma} \sum_{n,m=1}^{2} \frac{e^{i(\omega_n+\omega_m)t}}{-\omega_n\,\omega_m} \mathcal{X}_{\alpha\beta\gamma}(\boldsymbol{0}, \boldsymbol{0}; \omega_n, \omega_m) E_{n,\beta} E_{m,\gamma}. \tag{A9}$$

Note that $X_{\alpha\beta\gamma}(\boldsymbol{r}_1, \boldsymbol{r}_2; t_1, t_2)$ is a real-valued function, such that $\mathcal{X}_{\alpha\beta\gamma}(\boldsymbol{0}, \boldsymbol{0}; \omega_n, \omega_m) = \mathcal{X}_{\alpha\beta\gamma}(\boldsymbol{0}, \boldsymbol{0}; -\omega_n, -\omega_m)^*$. In addition, in the low-energy regime of a Weyl semimetal described by linearly dispersing bands, the corresponding current-density operator is associated with the zeroth-order term $\hat{j}_\alpha^{(0)}$ in the series expansion in Eq. (A2). Thus, contributions related to the operators $\hat{j}_\alpha^{(1)}$ and $\hat{j}_\alpha^{(2)}$ are vanished. Since the CPGE is associated with a current density generated by the term $\langle [\hat{j}_\alpha^{(0)}, [\hat{H}_{\text{pert}}^{(1)}, \hat{H}_{\text{pert}}^{(1)}]] \rangle_0$, we thus expect $\langle j_\alpha(\boldsymbol{r}, t) \rangle^{(\text{quad})}$ to contain the former observable as a nonvanishing contribution to the overall second-order response.

## APPENDIX B: INTERMEDIATE RESULTS FOR THE DRUDE-LIKE FORM OF THE TWO-POINT RETARDED GREEN'S FUNCTION

The starting point is the form of the three-current correlation function $\widetilde{\mathcal{X}}_{\alpha\beta\gamma}(i\Omega_1, i\Omega_2)$ given by Eq. (25) in the absence of internode scattering, after having performed the *ad hoc* approximations shown in Eqs. (26) and (27). Upon evaluating the frequency integrals exactly, and applying the recipe for the analytic continuation of the bosonic Matsubara frequencies mentioned in Sec. V, we obtain the following expression for the two-point retarded Green's function:

$$\mathcal{X}^{(\chi,\mathrm{SCBA,Drude})}_{\alpha\beta\gamma}(\omega_1,\omega_2)|_{\widetilde{W}_{\mathrm{inter}}=0}$$
$$\approx \frac{i}{12\pi^2}\chi\epsilon^{\alpha\beta\gamma}(-e)^3 v_F^3\int_0^{\Lambda/v_F} dk k^2 \sum_{b_1,b_2,b_3=\pm1}(1-\delta_{b_1b_2}\delta_{b_1b_3})$$
$$\times\Bigg(\frac{n_F(b_3v_Fk+|\mu_\chi|)}{\omega_1+\omega_2+(b_3-b_1)v_Fk+i\Gamma}\frac{1}{\omega_2+(b_3-b_2)v_Fk+i\Gamma}+\frac{n_F(b_2v_Fk+|\mu_\chi|)}{\omega_1+(b_2-b_1)v_Fk+i\Gamma}\frac{1}{-\omega_2+(b_2-b_3)v_Fk-i\Gamma}$$
$$+\frac{n_F(b_1v_Fk+|\mu_\chi|)}{-\omega_1+(b_1-b_2)v_Fk-i\Gamma}\frac{1}{-\omega_1-\omega_2+(b_1-b_3)v_Fk-i\Gamma}\Bigg)+(\beta\leftrightarrow\gamma,\omega_1\leftrightarrow\omega_2). \tag{B1}$$

The last term in Eq. (B1) is equal to the first one with the indices swapped as indicated. Taking the zero-temperature limit and rescaling the amplitude of momentum and energy UV cutoff as $Q=2v_Fk$ and $\widetilde{\Lambda}=2\Lambda$, respectively, we perform the summation over the parameters $b_j$ to find

$$\mathcal{X}^{(\chi,\mathrm{SCBA,Drude})}_{\alpha\beta\gamma}(\omega_1,\omega_2)|_{\widetilde{W}_{\mathrm{inter}}=0}\approx\frac{i}{96\pi^2}\chi\epsilon^{\alpha\beta\gamma}(-e)^3\int_{2|\mu_\chi|}^{\widetilde{\Lambda}} dQ\,Q^2\Bigg(\frac{1}{Q-(\omega_1+\omega_2+i\Gamma)}\frac{1}{Q-(\omega_2+i\Gamma)}$$
$$-\left[1-\frac{i\Gamma}{\omega_1+i\Gamma}\right]\frac{1}{Q+(\omega_1+\omega_2+i\Gamma)}\frac{1}{Q+(\omega_2+i\Gamma)}+\frac{1}{Q-(\omega_1+i\Gamma)}\frac{1}{Q+(\omega_2+i\Gamma)}$$
$$-\left[1+\frac{i\Gamma}{\omega_1+\omega_2+i\Gamma}\right]\frac{1}{Q+(\omega_1+i\Gamma)}\frac{1}{Q-(\omega_2+i\Gamma)}+\frac{1}{Q+(\omega_1+\omega_2+i\Gamma)}\frac{1}{Q-(\omega_1+i\Gamma)}$$
$$-\left[1-\frac{i\Gamma}{\omega_2+i\Gamma}\right]\frac{1}{Q-(\omega_1+\omega_2+i\Gamma)}\frac{1}{Q-(\omega_1+i\Gamma)}\Bigg)+(\beta\leftrightarrow\gamma,\omega_1\leftrightarrow\omega_2). \tag{B2}$$

Since we have assumed in Sec. V that $\Gamma\ll|\omega_1|,|\omega_2|,|\omega_1+\omega_2|$, we may drop the fractions in the rectangular brackets in Eq. (B2). In addition, we take the energy UV cutoff sufficiently large such that $\widetilde{\Lambda}\gg|\omega_1|,|\omega_2|,|\omega_1+\omega_2|$. Consequently, we find

$$\mathcal{X}^{(\chi,\mathrm{SCBA,Drude})}_{\alpha\beta\gamma}(\omega_1,\omega_2)|_{\widetilde{W}_{\mathrm{inter}}=0}$$
$$\approx\frac{i}{96\pi^2}\chi\epsilon^{\alpha\beta\gamma}(-e)^3\Bigg(\frac{(\omega_1+\omega_2+i\Gamma)^2}{\omega_2}\ln(4|\mu_\chi|^2-(\omega_1+\omega_2+i\Gamma)^2)-\frac{(\omega_1+\omega_2+i\Gamma)^2}{\omega_1}\ln(4|\mu_\chi|^2-(\omega_1+\omega_2+i\Gamma)^2)$$
$$+\frac{(\omega_2+i\Gamma)^2}{\omega_1+\omega_2+2i\Gamma}\ln(4|\mu_\chi|^2-(\omega_2+i\Gamma)^2)-\frac{(\omega_1+i\Gamma)^2}{\omega_1+\omega_2+2i\Gamma}\ln(4|\mu_\chi|^2-(\omega_1+i\Gamma)^2)$$
$$+\frac{(\omega_2+i\Gamma)^2}{\omega_1}\ln(4|\mu_\chi|^2-(\omega_2+i\Gamma)^2)-\frac{(\omega_1+i\Gamma)^2}{\omega_2}\ln(4|\mu_\chi|^2-(\omega_1+i\Gamma)^2)\Bigg)+(\beta\leftrightarrow\gamma,\omega_1\leftrightarrow\omega_2). \tag{B3}$$

By carefully inspecting the last term in Eq. (B3), we deduce that it is equal to the first term. We further choose the values for the real frequencies $\omega_1=\omega_0$, $\omega_2=-\omega_0+\delta\omega$, with optical frequency $\omega_0>0$ and frequency detuning $\delta\omega>0$ under the condition $\delta\omega\ll\omega_0$. Recalling the assumption $\Gamma\ll\delta\omega$ as well, we find that Eq. (B3) is dominated by terms $\propto(\omega_1+\omega_2+2i\Gamma)^{-1}$. Using the approximate form $\omega_0\pm i\Gamma\approx\omega_0$ in the prefactors of the logarithmic functions, the Drude-like two-point retarded Green's function finally reads

$$\mathcal{X}^{(\chi,\mathrm{SCBA,Drude})}_{\alpha\beta\gamma}(\omega_0,-\omega_0+\delta\omega)|_{\widetilde{W}_{\mathrm{inter}}=0}\approx\chi\frac{(-e)^3}{24\pi}\epsilon^{\alpha\beta\gamma}\frac{\omega_0^2}{\delta\omega+2i\Gamma}\frac{1}{\pi}\mathrm{Im}[\ln(4|\mu_\chi|^2-(\omega_0+i\Gamma)^2)]. \tag{B4}$$

The latter expression is shown in Eq. (29) in the main text, and it has been used in deriving the form of the Drude-like second-order current density in Eq. (31).